\documentclass[floats,floatfix,showpacs,amssymb,prd,twocolumn,superscriptaddress,nofootinbib,nolongbibliography,reprint]{revtex4-1}

\usepackage{amssymb,amsmath,verbatim,mathtools,needspace,enumitem,etoolbox,graphicx,physics,microtype,afterpage,xspace,tabularx,lmodern,multirow}
\usepackage{gensymb}
\usepackage[dvipsnames, usenames]{xcolor}
\definecolor{linkcolor}{rgb}{0.0,0.3,0.5}
\usepackage[unicode, colorlinks=true, linkcolor=linkcolor, citecolor=linkcolor, filecolor=linkcolor, urlcolor=linkcolor, linktocpage, breaklinks]{hyperref}
\usepackage[all]{hypcap}
\usepackage[T1]{fontenc}
\usepackage[utf8]{inputenc}
\usepackage[usenames,dvipsnames]{xcolor}
\hypersetup{colorlinks=true,citecolor=romared,linkcolor=romared,urlcolor=romared}

\definecolor{romared}{RGB}{142,0,28}

\newcommand{\be}{\begin{equation}}
\newcommand{\ee}{\end{equation}}

\def\be{\begin{equation}}
\def\ee{\end{equation}}
\newcommand{\beq}{\begin{eqnarray}}
\newcommand{\eeq}{\end{eqnarray}}

\usepackage{makecell}
\usepackage{soul}

\renewcommand{\eqref}[1]{Eq.~(\ref{#1})}
\newcommand{\Beq}{\begin{eqnarray}}
\newcommand{\Eeq}{\end{eqnarray}}

\def\lsim{\mathrel {\vcenter {\baselineskip 0pt \kern 0pt \hbox{$<$} \kern 0pt \hbox{$\sim$} }}}

\begin{document}

\title{Superradiance in massive vector fields with spatially varying mass}

\author{Zipeng Wang}
\email{zwang264@jhu.edu}
\affiliation{Department of Physics and Astronomy, Johns Hopkins University, Baltimore, MD 21218, USA}

\author{Thomas Helfer}
\email{thomashelfer@live.de}
\affiliation{Department of Physics and Astronomy, Johns Hopkins University, Baltimore, MD 21218, USA}

\author{Katy Clough}
\email{k.clough@qmul.ac.uk}
\affiliation{School of Mathematical Sciences, Queen Mary University of London Mile End Road, London, E1 4NS, UK}

\author{Emanuele Berti}
\email{berti@jhu.edu}
\affiliation{Department of Physics and Astronomy, Johns Hopkins University, Baltimore, MD 21218, USA}

\begin{abstract}
Superradiance is a process by which massive bosonic particles can extract energy from spinning black holes, leading to the build up of a condensate if the particle has a Compton wavelength comparable to the black hole's Schwarzschild radius. 
One interesting possibility is that superradiance may occur for photons in a diffuse plasma, where they gain a small effective mass.  Studies of the spin-0 case have indicated that such a build up is suppressed by a spatially varying effective mass, supposed to mimic the photons' interaction with a physically realistic plasma density profile.  We carry out relativistic simulations of a massive Proca field evolving on a Kerr background, with modifications to account for the spatially varying effective mass. This allows us to treat the spin-1 case directly relevant to photons, and to study the effect of thinner disk profiles in the plasma. We 
find similar qualitative results to the scalar case, and so support the conclusions of that work: either a constant asymptotic mass or a shell-like plasma structure is required for superradiant growth to occur. 
We study thin disks and find a leakage of the bosonic condensate that suppresses its growth, concluding that thick disks are more likely to support the instability.
\end{abstract}

\maketitle

\section{Introduction} \label{sect:intro}

In the presence of a highly spinning black hole (BH), massive bosonic fields can develop superradiant instabilities. The field can scatter off the BH in a way that extracts energy and angular momentum from it, in a wave analog to the Penrose process~\cite{Penrose:1971uk}. The fluctuations may be seeded by an initial environment or quantum fluctuations.  If the amplified excitations in the field cannot escape to infinity, they may fall back onto the BH such that the process will continuously repeat, forming a bosonic condensate with energy that grows exponentially over time~\cite{Cardoso:2004nk}.  The phenomenon of superradiance, first proposed by Zel'dovich~\cite{1971JETPL..14..180Z}, has been extensively studied using both numerical~\cite{Witek:2012tr,East:2013mfa,Okawa:2014nda,Zilhao:2015tya,East:2017ovw,East:2017mrj,East:2018glu} and (semi)analytical methods~\cite{Press:1972zz,Deruelle:1974zy,Damour:1976kh,Detweiler:1980uk,Gaina:1992nx,Zouros:1979iw,Lasenby:2002mc,Cardoso:2005vk,Dolan:2007mj,Grain:2007gn,Arvanitaki:2009fg,Arvanitaki:2010sy,Dolan:2012yt,Yoshino:2013ofa, Brito:2014wla,Arvanitaki:2014wva,Arvanitaki:2016qwi,Cardoso:2018tly,Frolov:2018ezx,Dolan:2018dqv,Ficarra:2018rfu,Siemonsen:2019ebd,Creci:2020mfg}: see~\cite{Brito:2015oca} for a comprehensive review.

The mass term naturally confines fields around BHs, producing a potential well such that they are gravitationally bound. In appropriate mass ranges, the extraction of energy and angular momentum continues for successive bound modes, until the spin of the black hole becomes too small to support further growth of the bosonic field. The saturation of each mode has been shown to occur via a smooth and approximately adiabatic process~\cite{East:2017ovw}. 

Beyond the idealized case of a pure Kerr background metric and a simple mass term in the bosonic potential, there are many physical mechanisms that may disrupt superradiance. A key outstanding question is how robust the process is to environmental effects and additional interactions. Even for a mass term in the absence of self-interactions, the presence of multiple modes can significantly affect the superradiant growth~\cite{Ficarra:2018rfu}. Self-interactions may lead to an early saturation of the superradiant instability due to mode mixing~\cite{Baryakhtar:2020gao} (although this requires further investigation in the relativistic regime~\cite{Omiya:2020vji}), with the possibility of an explosive destabilization of the condensate in a ``bosenova'' event~\cite{Yoshino:2012kn,Yoshino:2015nsa}. Recent studies have also considered how couplings to Standard Model particles may produce electromagnetic counterparts or quench the mechanism~\cite{Sen:2018cjt,Blas:2020nbs,Blas:2020kaa,Caputo:2021efm}, as well as the effect of deviations from the Kerr metric~\cite{Franzin:2021kvj,Guo:2021xao}.
  
\begin{figure*}[t]
  \includegraphics[width=1\textwidth]{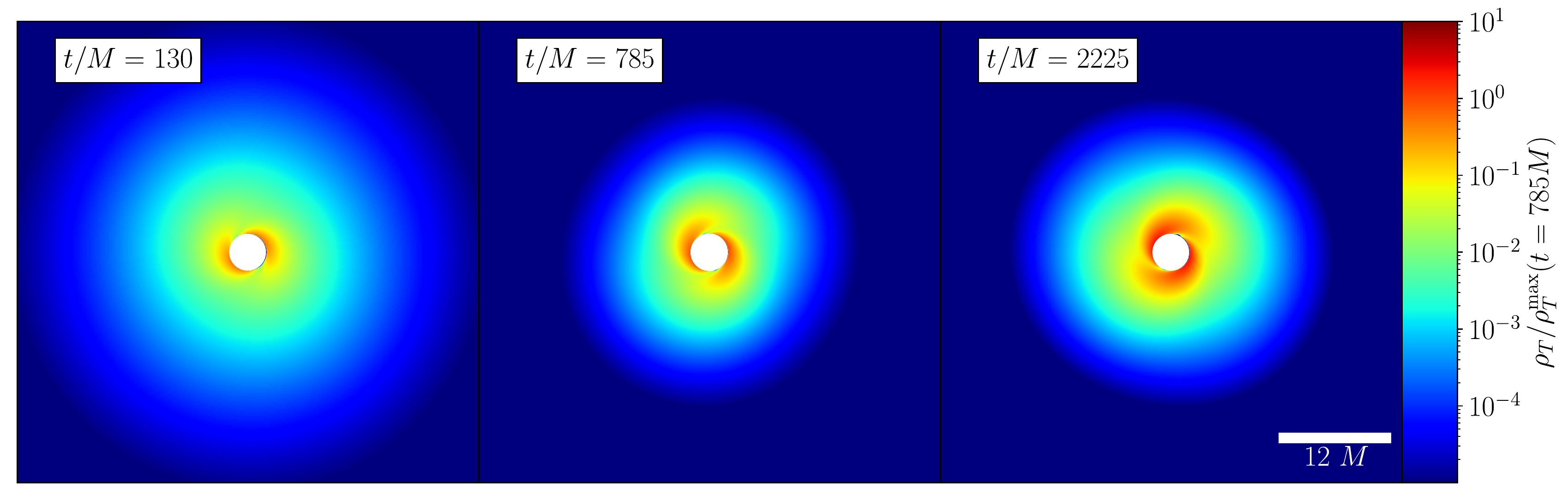}
  \caption{Time evolution in the $x-y$ plane of the Proca field energy density $\rho_T(t)$, as defined in \eqref{eqn:rho} below, for constant Proca mass $\mu=0.5M^{-1}$. We show snapshots at three different time steps: at $t=130M$, when the field is in the transient phase (i.e. nonsuperradiant modes in the initial data are decaying into the BH or radiating away); at $t=785M$, when the superradiant mode is just becoming dominant; and at $t=2225M$, when the Proca field has been growing in the superradiant phase for some time. The energy density is normalized to its maximum value $\rho_{T,{\rm ref}}$ for the simulation with $t=785M$. The values within the BH horizon are set to zero to mask the excision region.}
  \label{fig:constpic}
\end{figure*}

The superradiant process is highly dependent on the mass $M$ of the BH and the scalar field mass $m = \hbar \mu$, and is most efficient when the particle Compton wavelength $\lambda = 1/\mu$ is of the same order as the BH radius $r_s \sim M$ (throughout the paper we adopt geometrical units such that $G=c=1$).
In physical units this gives
\begin{equation}
    \mu M \simeq \left( \frac{M}{M_{\odot}}\right) \left(  \frac{m}{10^{-10} \text{eV} }\right) = \mathcal{O}(1)~, \label{eq:SRcondition}
\end{equation}
which sets a range of particle masses for solar mass and supermassive BHs for which the process may occur. As can be seen, these masses are much lower than those of any known bosonic particles, and so typically superradiance is of interest for light vector or scalar bosons beyond the Standard Model, e.g. dark photons or axion-like-particles, which may compose some fraction of the dark matter (but not necessarily a substantial amount).

However, it has also been proposed that superradiant instabilities might occur when the photon gains an effective mass while passing through a cloud of plasma~\cite{Cardoso:2005vk,Conlon:2017hhi,Dima:2020rzg}: for example, when the BH is immersed in a diffuse accretion disk. 
The effective mass of a photon in plasma is given by its oscillation frequency:
\begin{equation}
    \mu =  \sqrt{ 4 \pi \alpha \frac{n_e}{m_e} } = 1.2 \cdot 10^{-12}
    \sqrt{\frac{n_e}{10^{-3}~\text{cm}^{-3}}} \text{eV},
\end{equation}
where $n_e$ is the number density of the plasma~\cite{Dima:2020rzg}. For
stellar mass BHs ($M\sim 1-100~M_{\odot}$), the superradiant range in \eqref{eq:SRcondition} corresponds to $\mu \sim10^{-10} - 10^{-12}~\text{eV}$, and thus $n_e \sim 10^{-2} - 10^{-3} \text{cm}^{-1}$. 
Plasma densities around this range are typical in the interstellar medium (ISM)~\cite{Schnitzeler:2012jq}.
The considerations around the robustness of superradiance mentioned above are particularly important for this case, since treating the photon as a Proca field with a constant mass is clearly an oversimplification of the complicated magnetohydrodynamical effects involved, as was clearly acknowledged by those who originally proposed this mechanism~\cite{Conlon:2017hhi}. These less idealized configurations, including higher-order interactions in the plasma and electromagnetic fields~\cite{Conlon:2017hhi,Cardoso:2020nst,Blas:2020kaa,Cannizzaro:2020uap,Cannizzaro:2021zbp}, or spatially varying densities~\cite{Cardoso:2013opa,Dima:2020rzg}, may mean that the build up does not occur, or that if it does the growth may end in runaway instabilities like the bosenova~\cite{Yoshino:2012kn,Yoshino:2015nsa,Yoshino:2013ofa,Baryakhtar:2020gao,Yoshino:2015nsa,Witek:2012tr,Arvanitaki:2010sy}.

There are several interesting consequences, should plasma-driven superradiance indeed be effective and robust. It may be used to explain the low BH spins measured by the Advanced LIGO and Virgo network of gravitational-wave detectors~\cite{Dima:2020rzg}, and increase the fraction of hierarchical mergers in dynamical formation scenarios~\cite{Payne:2021ahy}. It would also potentially prevent BH spin observations putting constraints on the existence of new bosonic particles, as first proposed in~\cite{Arvanitaki:2009fg} (see also~\cite{Arvanitaki:2010sy,Brito:2014wla,Arvanitaki:2014wva,Yoshino:2014wwa,Arvanitaki:2016qwi,Baryakhtar:2017ngi,Ng:2020ruv}), and complicate the search for a stochastic background of gravitational waves~\cite{Brito:2017wnc, Brito:2017zvb, Siemonsen:2019ebd, Tsukada:2020lgt, Zhu:2020tht}. The impact of the effect has also been investigated in the context of compact stars~\cite{Cardoso:2017kgn,Day:2019bbh}.
A further potential application of the effect is in the early Universe, when plasma densities were higher and thus the corresponding photon mass was much larger. 
Then a superradiant instability could be triggered around light primordial black holes~\cite{Pani:2013hpa}, resulting in a spectral distortion of the CMB. Since we do not observe such a distortion, one could place bounds on the existence of highly spinning PBHs, provided the mechanism is considered sufficiently robust to occur in a generic situation.  Finally, it has been speculated that destabilization of the superradiant build up in plasmas may explain energetic transient signals at radio frequencies, such as Fast Radio Bursts (FRBs)~\cite{Lorimer:2007qn,Katz:2016dti,Conlon:2017hhi,Houde:2018yos}.

Dima et al.~\cite{Dima:2020rzg} studied the effect of a position-dependent effective mass on the superradiant build up as a toy model for more realistic scenarios where the ISM plasma forms an accretion disk around the BH, whose density typically increases as it approaches the horizon. They used time-domain studies of a massive scalar field, using the spectral decomposition method of Dolan~\cite{Dolan:2013} to follow the long timescales involved.
In this work we perform similar studies in the massive vector boson (Proca) case, that is more directly applicable to photons. Vector bosons are more efficient at extracting energy and angular momentum from rotating BHs than scalars since their scattering results in a higher amplification of the incoming waves, and the superradiant modes are more closely bound, i.e., they are located nearer the BH horizon~\cite{Brito:2015oca}. One might hope that this could improve the robustness of the process to spatial variations in the plasma.
To study vector boson superradiance, we use a (3+1) dimensional evolution of the vector field on a stationary Kerr background (neglecting backreaction of the field on the metric). To facilitate comparison, we focus on effective mass configurations that are qualitatively similar to those in~\cite{Dima:2020rzg}, but our setup allows for more general mass configurations, so we can also probe the thin disk case. A typical evolution for the constant-mass case is shown in Fig.~\ref{fig:constpic}.

\begin{figure*}
  \includegraphics[width=0.95\textwidth]{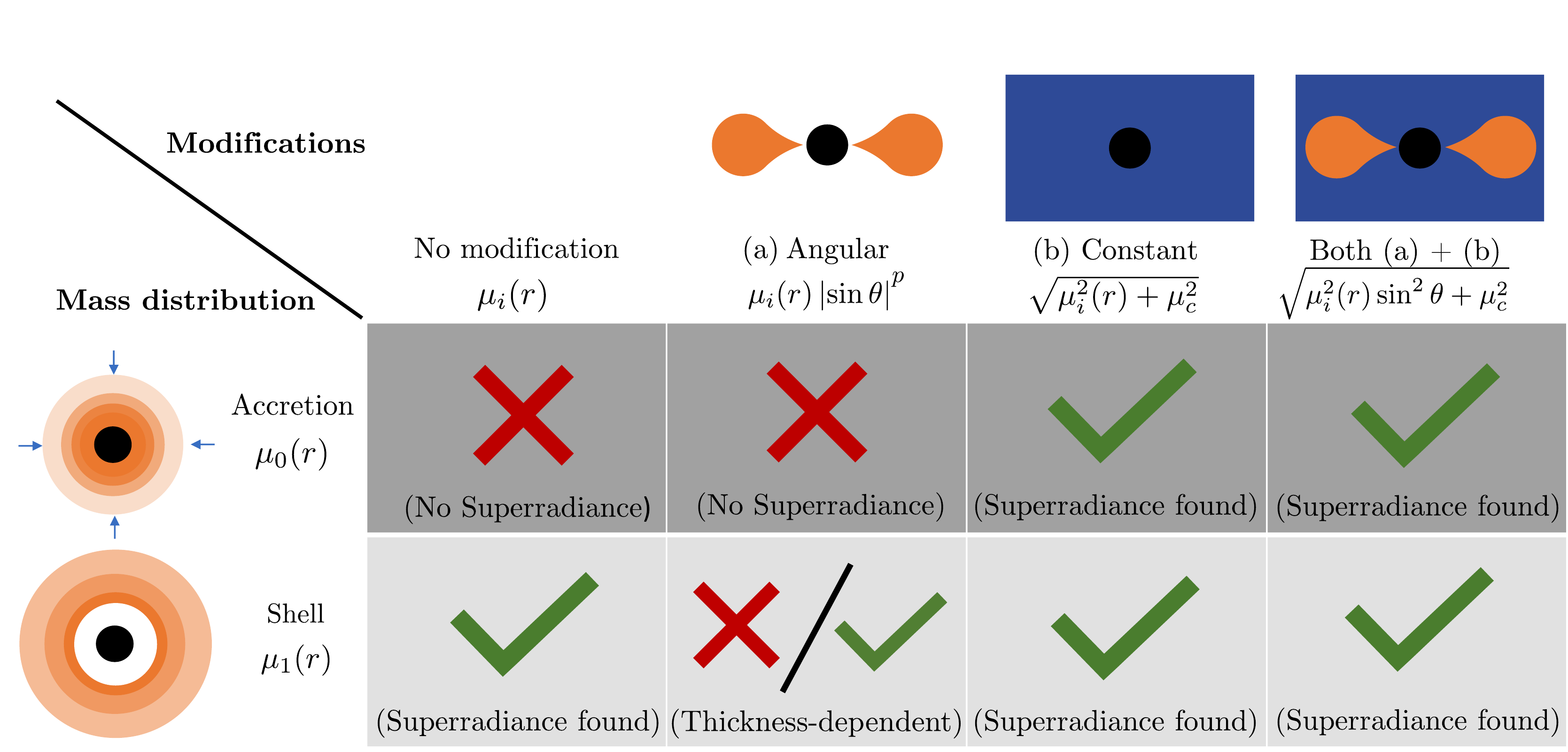}
  \caption{Schematic overview of the Proca mass profiles studied in this paper. In all cases we considered a BH spin parameter $a/M=0.99$. Rows correspond to the two radial mass profiles we study, as defined in Eqs.~(\ref{eq:mu0}) and (\ref{eq:mu1}) below: a ``spherical accretion'' profile $\mu_0(r)$ and a ``hollow shell'' profile $\mu_1(r)$. The columns show the effect of various modifications to these profiles, including the addition of an angular dependence (colored in orange in the schematic graphs on the top row) and/or of a nonzero asymptotic mass value (colored in blue). A red cross means that we did not observe modes that undergo superradiant growth for the simulated parameters, and a green tick mark means that we did. In one case (model 6 in the text), superradiant amplication may or may not occur depending on the shell's thickness.}
    \label{fig:schema}
\end{figure*}

Our results broadly support the conclusions reached by Dima et al.~\cite{Dima:2020rzg}. We find that superradiance does not occur when the effective mass background corresponds to a Bondi accretion profile. However, superradiant growth is once again possible with suitable modifications to the configuration, such as the addition of a constant asymptotic mass or a cut-off in the density in the inner region (creating a plasma ``shell'').  Compared to the scalar case in Ref.~\cite{Dima:2020rzg}, the enhancements we observe are more modest: our simulations do not yield order-of-magnitude improvements in the instability growth rates with respect to the constant-mass case. As above, our setup also permits the study of a wider range of disk thicknesses. In the absence of an asymptotic mass, we find that thin disks do not support superradiant growth (even with a sharp inner cut-off in the density), due to ``leakage'' of the bosonic field out of the poles. Therefore, all other factors being equal, thicker disks are more effective at triggering superradiance.
A schematic summary of our results is provided in Fig.~\ref{fig:schema}.

The paper is structured as follows. In Sec. \ref{sec:setup} we describe the numerical setup, with further technical details contained in Appendix~\ref{app:numerical}. In Sec. \ref{sec:results} we describe our results for the set of models tested. In Sec.~\ref{sec:discussion} we discuss our findings and future research directions.

\section{Setup and numerical methods}
\label{sec:setup}

\subsection{Kerr metric background}
\label{sec:kerr}

Following Refs.~\cite{East:2017ovw, East:2017mrj, Witek:2012tr}, we write the fixed background metric in Cartesian Kerr-Schild (KS) coordinates.  These have the advantage of being horizon penetrating, such that there is no coordinate singularity at the horizon (see~\cite{Visser:2007fj} for a discussion of this form of the Kerr metric and its interpretation). In these coordinates, the spacetime line element is given by:
\begin{eqnarray}
    ds^2 = -  dt^2 +dx^2 + dy^2 + dz^2 + \frac{2Mr^3}{r^4 + a^2z^2} \times \nonumber \\
     \left[ dt + \frac{r(x\, dx + y\, dy)}{a^2 +
    r^2} +   \frac{a(y\, dx - x\, dy)}{a^2 + r^2} +  \frac{(z~dz)}{r}  \right]^2\,.
  \label{KerrKS}
\end{eqnarray}
Here $r$ is not a coordinate (in particular, it is not the Cartesian radial coordinate $R = \sqrt{x^2 + y^2 + z^2}$) but rather a function of the Cartesian spatial coordinates  ($x$, $y$, $z$) given by the implicit expression
\begin{equation}
	\frac{x^2 + y^2}{r^2 + a^2} + \frac{z^2}{r^2} =  1\,,
\end{equation}
although one may recognize it as the Boyer-Lindquist radial coordinate from that alternative coordinate choice.  Here $M$ and $J$ denote the mass and angular momentum of the BH, $a=J/M$ is the Kerr parameter, and $a/M \in [0,1]$ is the dimensionless spin parameter.  Thus the BH is entirely parametrized by $J$ and $M$.  The alignment of the angular momentum is taken to be in the $z$ direction, without loss of generality.

\begin{figure*}[ht]
  \includegraphics[width=1\textwidth]{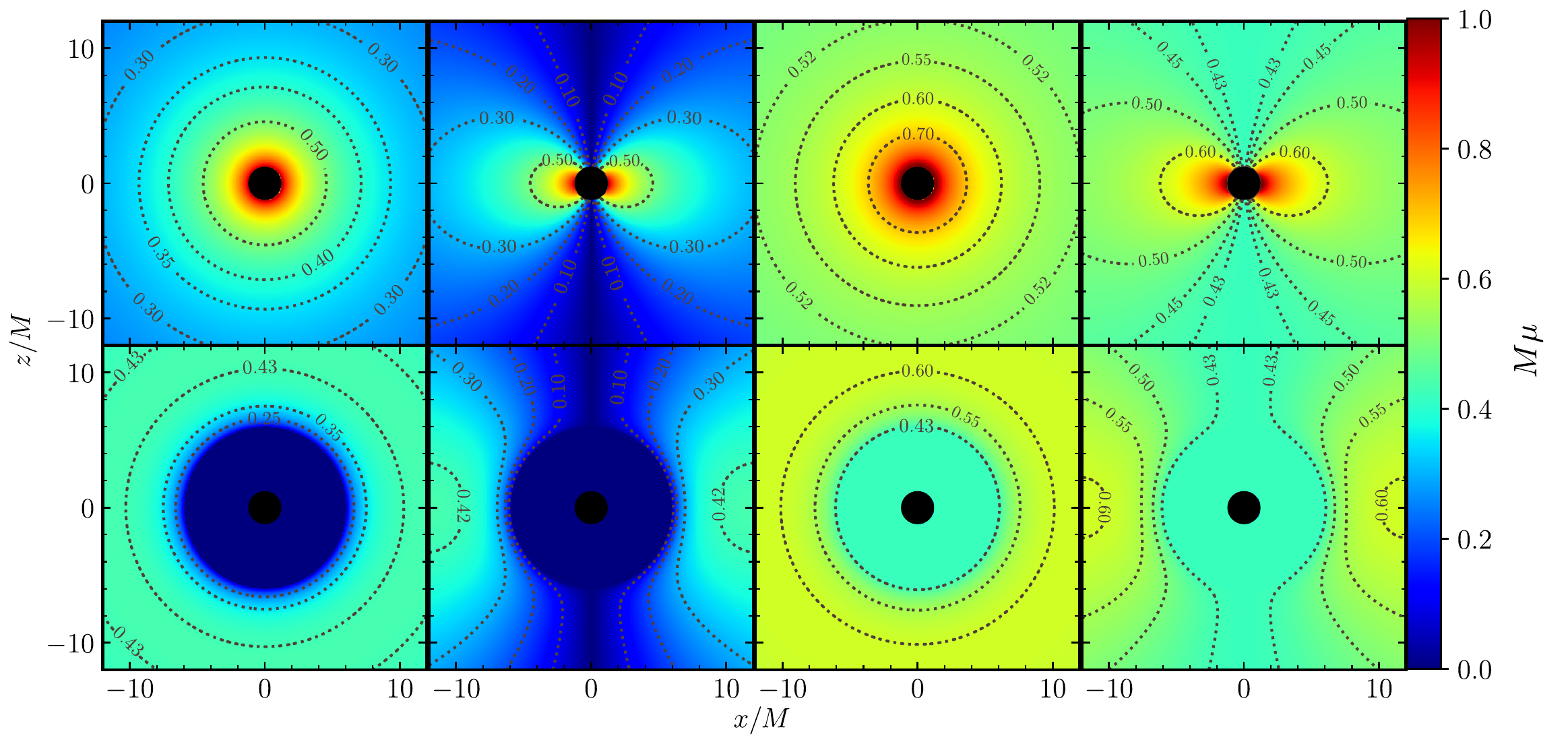}
  \caption{Proca mass profiles in Table \ref{table:mass} around the BH in the $x-z$ plane. The ordering of the panels shown here is consistent with that in Fig. \ref{fig:schema} and Fig. \ref{fig:multipanel}. The top row is generated using parameters $\mu_H = 1$, $\lambda=1$, $\mu_c=0.42$, and the bottom row is generated using $M \mu_H = 2$, $\lambda=1$, $M\mu_c=0.42$, and $r_0/M=6$, for illustration purposes. The central black circle masks the excision region around the inner horizon.}
  \label{fig:mass_profile}
\end{figure*}

The metric in the standard  $3+1$ ADM decomposition is given by:
\begin{equation}
ds^2=-\alpha^2\,dt^2+\gamma_{ij}(dx^i + \beta^i\,dt)(dx^j + \beta^j\,dt)\,,
\end{equation}
with components
\begin{equation}
	\alpha = (1 + 2H l^t l^t)^{-1/2}\,,
\end{equation}
\begin{equation}
	\beta^i = - \frac{2H l_t l_i}{1 + 2 H l_t l_t}\,.
\end{equation}
The induced metric reads
\begin{equation}
	\gamma_{ij}dx^i dx^j = \delta_{ij} + 2H l_i l_j\,,
\end{equation}
where
\begin{equation}
	H = \frac{Mr^3}{r^4 + a^2 z^2}\,,
\end{equation}
\begin{equation}
	l_\mu = \left( 1, \frac{rx + ay}{r^2 + a^2},  \frac{ry - ax}{r^2 + a^2},
    \frac{z}{r}  \right)\,.
\end{equation}

The extrinsic curvature $K_{ij}$ is derived from its definition as
\begin{equation}
	K_{ij} = \frac{1}{2 \alpha}(D_i\beta_j + D_j \beta_i)\,,
\end{equation}
given that $\partial_t \gamma_{ij} = 0$.

\subsection{Proca field}

The real Proca field $X^\mu$ is associated with the Lagrangian
\begin{equation}
	\mathcal{L} = \frac{1}{4} F_{\alpha\beta} F^{\alpha\beta} + \frac{1}{2}
    \mu^2 (X_\alpha X^\alpha)\,,
\end{equation}
and evolves on the background metric according to the equation of motion
\begin{equation}
	\nabla_{\beta}F^{\alpha\beta} = \mu^2 X^\alpha ~,  \label{eqn-proca_eom}
\end{equation}
where $F_{\alpha\beta}$ is defined as
\begin{equation}
F_{\alpha\beta} = \nabla_{\alpha}X_{\beta}-\nabla_{\beta} X_{\alpha}\,. 
\end{equation}

Assuming that the Proca field mass $\mu$ is a constant, this results in the requirement that
\begin{equation}
\nabla^{\alpha}X_{\alpha} = 0~. \label{eqn-lorenz}
\end{equation}
Note that since $X_\alpha$ is massive, this condition is not a gauge choice, but a constraint that must be satisfied.

In our toy model, the mass term $\mu$ depends on the spatial coordinates. In this case the constraint becomes
\begin{equation}
    \label{eqn-new_lorenz}
    \nabla_\alpha X^\alpha = - 2 ~ \partial_\alpha (\ln \mu) X^\alpha\,,
\end{equation}
which is still a coordinate-invariant expression.

We decompose \eqref{eqn-proca_eom} into (3+1) dimensional ADM coordinates, following Ref.~\cite{Zilhao:2015tya}. Using the projection operator of the spatial slices
\begin{equation}
P_{\mu}^{\nu} = \delta_{\mu}^{\nu}+n_{\mu}n^{\nu}\,,
\end{equation}
where $n^{\mu}$ is the normal to the hypersurface, the field $X_\mu$ can be
decomposed into a spatial part $A_i$ and a time-like part $\varphi$, where
\begin{equation}
A_\mu = P_{\mu}^{\nu} X_{\nu}   \quad   \text{and} \quad \varphi = - n^{\mu} X_{\mu} \,.
\end{equation}
An electric field is defined by analogy with electromagnetism, which provides
the equation for the (first-order in time) evolution of $A_i$
\begin{equation}
E_i= P_{i}^{\mu}n^{\nu}F_{\mu\nu} \,.
\end{equation}

Projection of the 4-dimensional equation of motion onto the spatial slice and normal to it gives rise to an equation of motion for $E_i$ and the constraint
\begin{equation}
	\mathcal{C_E} = D_i E^i - \mu^2 \varphi = 0\,, \label{eqn:constraint}
\end{equation}
where $D_i$ is the covariant derivative on the spatial slice. The evolution
equation for $\varphi$ is derived from \eqref{eqn-new_lorenz}.
To ensure that numerical violation of \eqref{eqn:constraint} is kept to a
minimum, we stabilize it by introducing an auxiliary damping variable $Z$~\cite{Zilhao:2015tya,Hilditch:2013sba,Palenzuela:2009hx}.
The equations of motion for the decomposed quantities in terms of the ADM metric variables are then
\begin{align} \label{eqn-ADM_phi}
\partial_t \varphi &= - A^i D_i \alpha + \alpha (K\varphi - D_i A^i - Z) + 
    \mathcal{L}_{\beta} \varphi   \nonumber \\
    &- 2 \left( \alpha A^i - \beta^i \varphi \right)  \partial_i (\ln \mu)\,,\\
\partial_t A_i &= - \alpha (E_i + D_i \varphi) - \varphi D_i \alpha +
    \mathcal{L}_{\beta} A_i\,, \\
\partial_t E^i &=  \alpha (K E^i + D^i Z + \mu^2 A^i + D^k D_i A_k - D_i D^k A_k ) \nonumber \\
                     &+ D^j \alpha (D^i A_j - D_j A^i))  +  \mathcal{L}_{\beta}
                     E_i\,, \\
\partial_t Z &= \alpha ( D_i E^i + \mu^2 \varphi - \kappa Z) + \mathcal{L}_{\beta} Z \,,
\end{align}
where $\mathcal{L}$ denotes the Lie derivative, and $\kappa$ is a constant of order unity that controls the level of constraint damping. We use Sommerfeld boundary conditions to allow outgoing radiation to exit the grid with minimal reflections.

\subsection{Initial Proca Data}

We follow the suggestion in Ref.~\cite{East:2017mrj} and use an initial ``seed''
for the superradiant growth of the lowest $m=1$ ($S=-1$, $n=0$) mode of the form
\begin{equation}\label{eqn:IC}
\begin{split}
    A_x &= A_y = \frac{A}{\det{(\gamma_{ij})}}  e^{-R / r_0}, \\
    A_z &= 0, \\
\end{split}
\end{equation}
where $R_0 \approx 1 / ( M \mu^2 )$ is approximately the characteristic radius for the $m=1$ mode, 
and $A \sim 10^{-1}$ is a small seed amplitude (the absolute value of $A$ is arbitrary when we neglect backreaction, as we do here).

\subsection{Diagnostic quantities}
\label{sec-diagnostics}

The stress-energy tensor for the Proca field is
\begin{equation}
T_{\mu\nu} = F_\mu^\rho F_{\nu \rho} - \frac{1}{4} g_{\mu\nu} F_{\rho\sigma} F^{\rho \sigma} + \mu^2 X_\mu X_\nu - \frac{\mu^2}{2} g_{\mu\nu} X_\rho X^\rho \,.
\end{equation}
We define the projections of the stress-energy tensor
\begin{equation}\label{eq:decomposed-EM-tensor}
    \rho \equiv n_{\alpha} n_{\beta} T^{\alpha\beta}\,,
    \quad
    S_i \equiv -\gamma_{i\alpha}n_{\beta} T^{\alpha\beta}\,,
    \quad
    S_{ij} \equiv \gamma_{i\alpha}\gamma_{j\beta} T^{\alpha\beta}\,,
\end{equation}
which we can calculate as 
\begin{equation}\label{eqn:energy}
\rho = \frac{1}{2} \left( E_i E^i + \mu^2 (\varphi^2 + A_i A^i) + D_i A_j (D_i
    A_j - D_j A_i) \right)\,,
\end{equation}
\begin{equation}
S_{i} = \mu^2 \varphi A_i + E^j (D_i A_j - D_j A_i) \,,
\end{equation}
and 
\begin{equation}
\begin{aligned}
    S_{ij} =& \mu^2 \left[ A_i A_j + \frac{1}2 \gamma_{ij} \left(
    \varphi^2-A^kA_k  \right) \right] + E_i E_j \\
    &+ \frac{1}2 \gamma_{ij} E^k E_k + (\partial_i A_k - \partial_k
    A_i)(\partial_j A^k - \partial_k A^j) \\
    &-\frac{1}2\gamma_{ij}D^l A^m (\partial_l A_m - \partial_m A_l) \,.
\end{aligned}
\end{equation}

Due to the existence of a time-like Killing vector $\zeta^\mu = (1,0,0,0)$, the current $J^{\mu} = \sqrt{-g} ~\zeta_{\nu} ~T^{\mu\nu} $ is conserved: $\nabla_\mu J^{\mu} = 0$.
This results in a conservation equation obeyed on each 3-dimensional hyperslice of the 3+1 ADM decomposition~\cite{East:2017mrj,Clough:2021qlv}:
\begin{equation}\label{eqn:conservation}
    \partial_t \int_{\Sigma} ~ \sqrt{\gamma} \rho_T = \int_{\partial\Sigma} d^2x
    \sqrt{\sigma} ~ F_T\,,
\end{equation}
where
\begin{equation}\label{eqn:rho}
\begin{split}
    \rho_T &= n_\nu J^\nu  =  \alpha \rho - \beta_k S^k\,,
\end{split}
\end{equation}
\begin{equation}
    \begin{split}\label{eqn:F}
    &F_T = \alpha N_i J^i  = N_i \left[\beta^i (\alpha \rho - \beta^j S_j  ) +
    \alpha \gamma^{ij} (\beta^k S_{jk} - \alpha S_j ) \right]\,,
\end{split}
\end{equation}
and $n^{\nu}$ is the normal to the 3-dimensional hypersurface. Here $N^i$ is the normal to the 2-dimensional surface enclosing the 3-dimensional volume over which the density is integrated, for which the induced metric $\sigma_{ij}$ has determinant $\sigma$. Further details on implementing these expressions in a simulation can be found in Ref.~\cite{Clough:2021qlv}.

In our simulations we monitor the individual contributions to \eqref{eqn:conservation}, with the fluxes taken through coordinate spheres of radius $144M$ far from the bound superradiant mode (to check outgoing radiation of the Proca field), and $2M$ close to the BH horizon (to monitor fluxes through the horizon). These fluxes are reconciled to the growth of mass-energy within the volume, to monitor the accumulation of errors in the simulation.  The overall growth of the mass of the bound Proca field state is used as a measure of the superradiant growth rate.

\subsection{Numerical methods}

In our simulations, the metric of \eqref{KerrKS} is a fixed background metric on which the dynamical Proca field evolves. Therefore we are neglecting backreaction of the Proca field on the metric, and calculating the mass profile $\mu$, the metric values and their gradients analytically at each point. This will be a good approximation where the energy density of the Proca field is small, as it would be at the initial stages of the superradiant build up in which we are interested. We fix the value of $a/M = 0.99$ and $M=1$ in code units.

The Kerr-Schild form of the metric necessitates excision of the singularity, which is achieved by setting the field components and their time derivatives to zero just outside the inner horizon.  Given sufficient resolution at the outer horizon, the ingoing nature of the metric prevents the errors this introduces from propagating to the region outside the BH.  For the value of $a=0.99M$ used in this work, the outer horizon is a spheroid with equatorial radius $1.511M$ and polar radius $1.141M$. We found that a finest resolution of $\Delta x= \frac{1}{48} M \approx 0.02083 M$ was required to recover the correct rate of superradiant growth in the constant-mass case, and the same value was found to achieve convergence in our simulations with a spatially varying mass.  The finest grid at this resolution covers a cube with side length $ \frac{10}3M = 3.33M$, which fully encloses the outer horizon.

Further numerical details, including code verification and convergence testing, are provided in Appendix \ref{app:numerical}.

\begin{table}
\begin{center}
\begin{tabular}{ l  l }
Models  & $\mu(x)$  \\
\hline
    1) Accretion & $ \mu_0(r) $  \\
    2) Accretion-angular & $\mu_0(r)\left|\sin \theta\right|$ \\
    3) Accretion-constant & $\sqrt{\mu_0(r)^2+\mu^2_c} $  \\
    4) Accretion-angular-constant & $\sqrt{\mu_0(r)^2\sin^2 \theta+\mu^2_c} $ \\
    5) Shell & $\mu_1(r)^2 $  \\
    6) Shell-angular & $\mu_1(r)\left |\sin \theta \right|^p $ \\
    7) Shell-constant & $\sqrt{\mu_1(r)^2+\mu_c^2} $ \\
    8) Shell-angular-constant & $\sqrt{\mu_1(r)^2\sin^2 \theta + \mu_c^2} $ \\
\hline
\end{tabular}
    \caption{{Proca mass profiles studied in this paper and in Ref.~\cite{Dima:2020rzg}}. Here $({r,\theta})$ are the radius and the polar angle in Boyer-Lindquist coordinates, $r_+$ is the outer horizon of the black hole, and $\mu_c$, $\mu_H$, $r_0$, $\lambda$ are constant parameters. The radial function $\mu_0(r)$ mimicking accretion is given in \eqref{eq:mu0}, and the function $\mu_1(r)$ mimicking a shell in \eqref{eq:mu1}.}
    \label{table:mass}
\end{center}
\end{table}

\begin{table*}[t]
    \begin{center}
      \begin{tabular}{ l  c c c  c  c  c c c}
        \hline
        Models  & \multicolumn{5}{c}{Parameters}
        &Growth/decay rate & Relative rate & Growth/decay rate\\        
                & $M\mu $ & $M\mu_H$ & $\lambda$  & $r_0/M$ & $p$ & $[10^{-4}/M]$ & & Dima et al. $[10^{-4}/M]$\\     
        \hline    
      ~\\
      0) Uniform & 0.5&--&--&--&--& $3.35\pm0.08$ & $1.00\pm0.02$ & --\\
        1) Accretion &-- &1.0 &1.5 &--&--  &$-61.0\pm4.2$ &
        $-18.2\pm1.2$ & $-4.99 \times 10^{2}$\\
        2) Accretion-angular &-- & 0.5 &2.5 & -- &1.0 &$-44.9\pm2.8 $
        & $-13.4\pm0.8$  &--\\
        3) Accretion-constant & 0.42 &0.5 &2.5 &--&--  &$1.23\pm0.02$ & $0.367\pm0.006$ & $3.99 \times 10^{-6}$\footnote{Ref.~\cite{Dima:2020rzg} used $M\mu_H = 1.0$ and $\lambda=2.0$, instead of the parameters listed in this Table.}\\
        4) Accretion-angular-constant & 0.42 &0.5 &2.5 &-- &--&$1.81\pm0.03$ &$0.539\pm0.010$ & --\\
        5) Shell &-- &2 & 2&8 &-- &$2.58\pm0.04$ & $0.768\pm0.011$ & $0.513$ \\
                 & -- &2 &2 &$r_{\text{ISCO}}$  &--&$-12.0\pm1.4$&
        $-3.58\pm0.40$ &$-2.62\times 10^{2}$ \\
        6) Shell-angular & --& 0.5& 2& 8 &1.0&$-7.19\pm0.05$ & $-2.12\pm0.02$ &--\\
                 & --& 2& 2& 8 &1.0 &$-5.58\pm0.04$ & $-0.17\pm0.01$&--\\
                 & --& 2& 2& 8 &0.75&$1.41\pm0.06$ & $0.42\pm0.02$ &--\\
                 & --& 2& 2& 8 &0.5&$1.54\pm0.07$ & $0.46\pm0.02$ &--\\                 
        7) Shell-constant & 0.42&2 &1.5 & 8  &--&$9.68\pm0.26$ &$1.44\pm0.04$ & --\\
        8) Shell-angular-constant & 0.42&2 &1.5 & 8 &-- &$8.69\pm0.18$ &$1.30\pm0.03$ & $-1.39$\footnote{Ref.~\cite{Dima:2020rzg} found a positive (growth) rate of $0.119$ when $M\mu_c=0.3$.} \\
      \hline      
    \end{tabular}
        \caption{Parameters used in each simulation, and resulting Proca field growth (plus sign) or decay (minus sign) rates. The first column refers to the mass profiles listed in Table~\ref{table:mass} (``Uniform'' denotes a constant Proca mass value). Columns 2-6 list the parameters used in each simulation. Column 7 lists the growth or decay rates obtained by an exponential fitting of the Proca energy data displayed in Fig.~\ref{fig:multipanel} for $t>1000M$, with errors estimated using fourth-order Richardson extrapolation (see Appendix~\ref{sec:richardson}). In column 8 we normalize the rate to the corresponding growth rate for a constant Proca mass (i.e., the ``Uniform'' case). In column 9, for comparison, we list the growth or decay rates found in Ref.~\cite{Dima:2020rzg}, when available.}
        \label{table:growth}
    \end{center}
\end{table*}

\begin{figure}[ht]
  \includegraphics[width=\columnwidth]{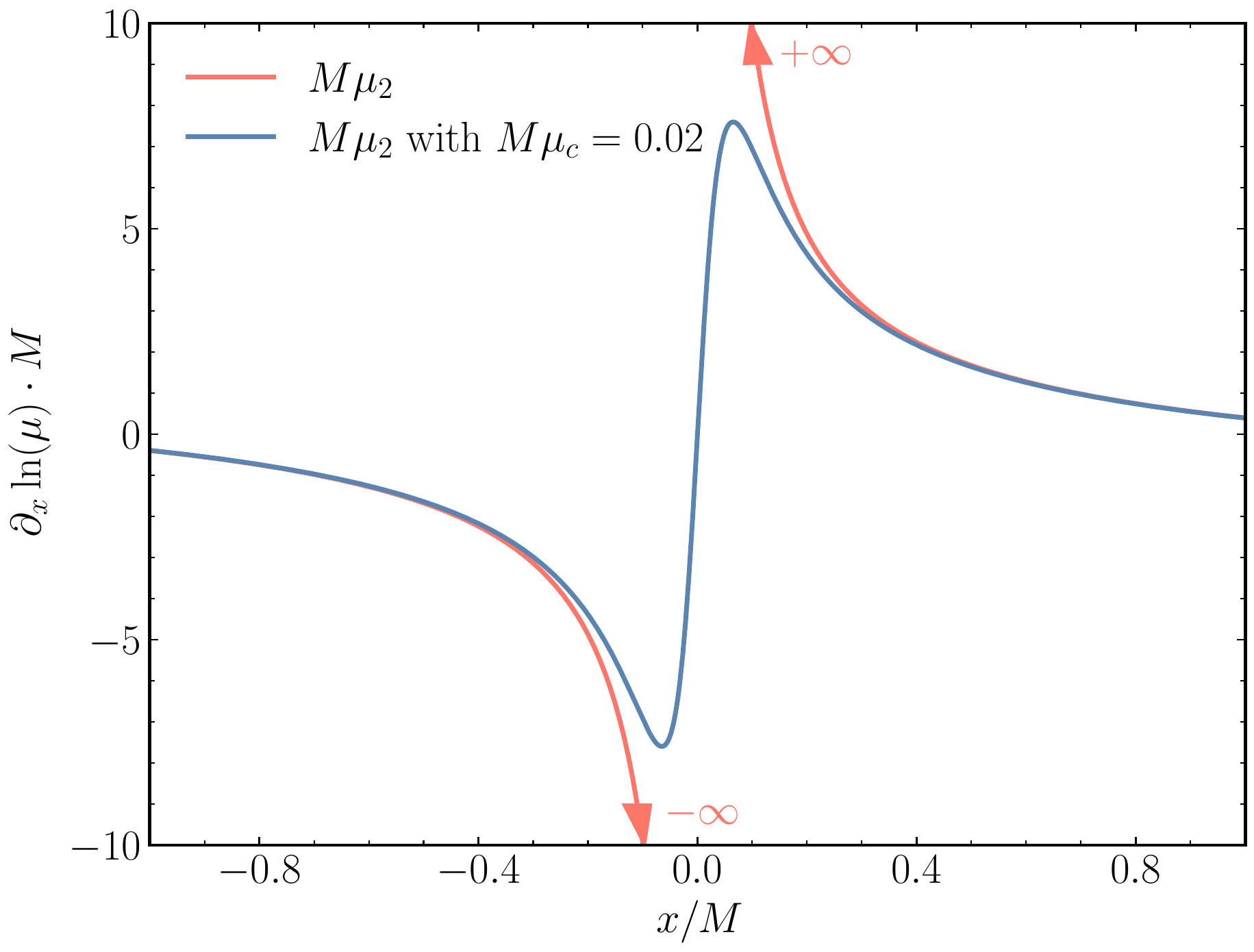}
    \caption{Derivative of $\ln\mu$, the natural logarithm of the Proca mass, for mass profiles with angular dependence. We plot the derivative of $\ln\mu$ with respect to the Kerr-Schild coordinate $x$ at $y=1.2M$, $z=0M$, and in the range $x \in [-M,\,M]$. The red curve refers to Model 2 (Accretion-angular) as defined in Table~\ref{table:mass} and Ref.~\cite{Dima:2020rzg}, and shows that this quantity diverges. The blue curve shows that the addition of a small constant mass term $\mu_c=0.02M^{-1}$ to Model 2 removes the divergence.}
    \label{fig:mu2deri}
\end{figure}

\begin{figure*}[t]
  \includegraphics[width=1\textwidth]{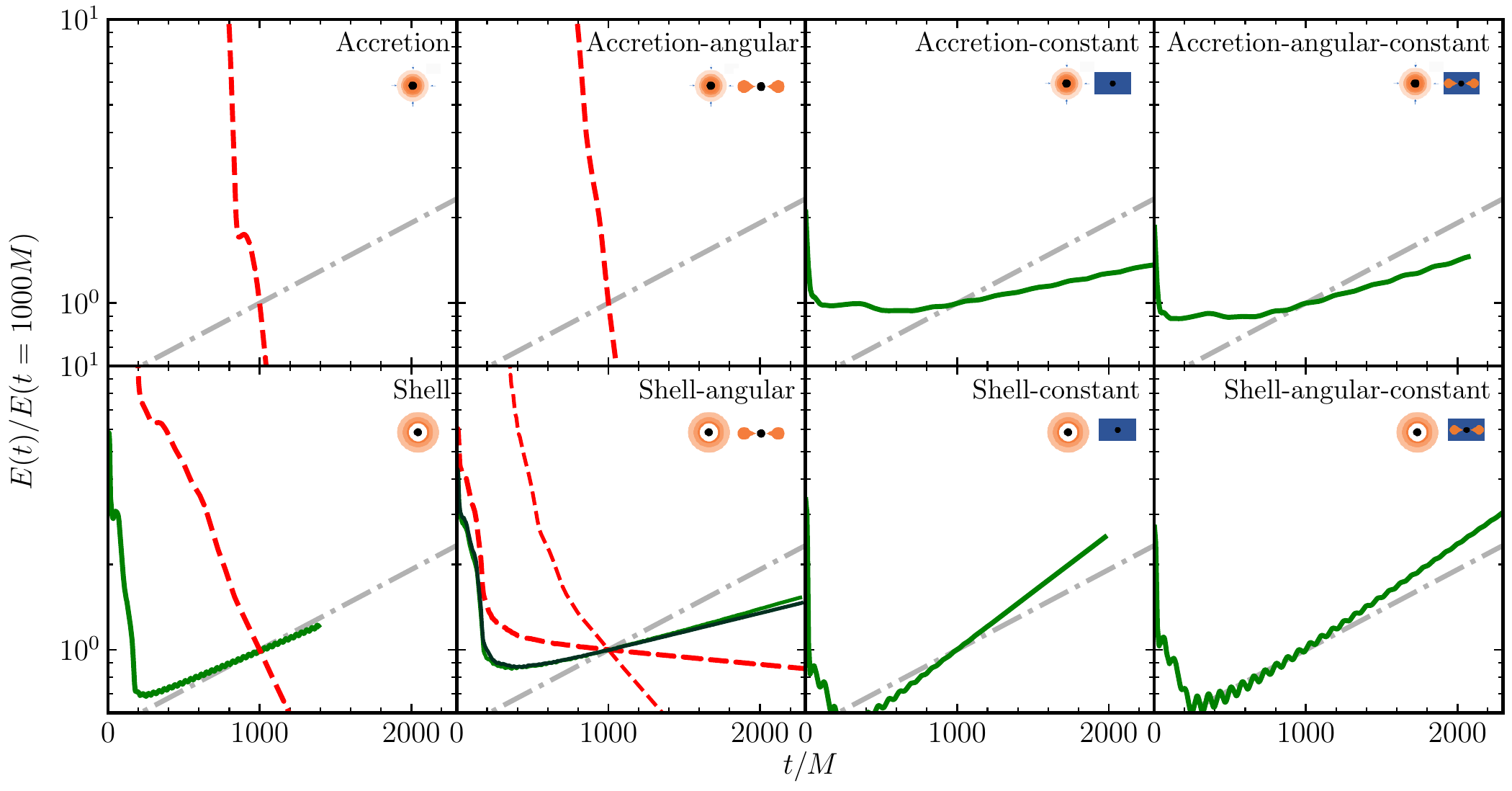}
  \caption{Time evolution of the Proca field energy $E(t)$, exhibiting either superradiant growth or decay for different Proca mass profiles. Each panel corresponds to the corresponding mass profile shown schematically in Fig.~\ref{fig:schema}. For reference, the gray dash-dotted lines in all eight panels show the simulation result for a constant Proca mass $M\mu=0.5$. All of the field energies $E(t)$ shown here are normalized to $E(t = 1000M)$ to aid comparison. The ``shell-angular'' case illustrates four profiles with varying radial dependence and different angular dependence (shell thickness), illustrating that thicker shells favor superradiance.}
  \label{fig:multipanel}
\end{figure*}

\begin{figure*}[ht]
  \includegraphics[width=1\textwidth]{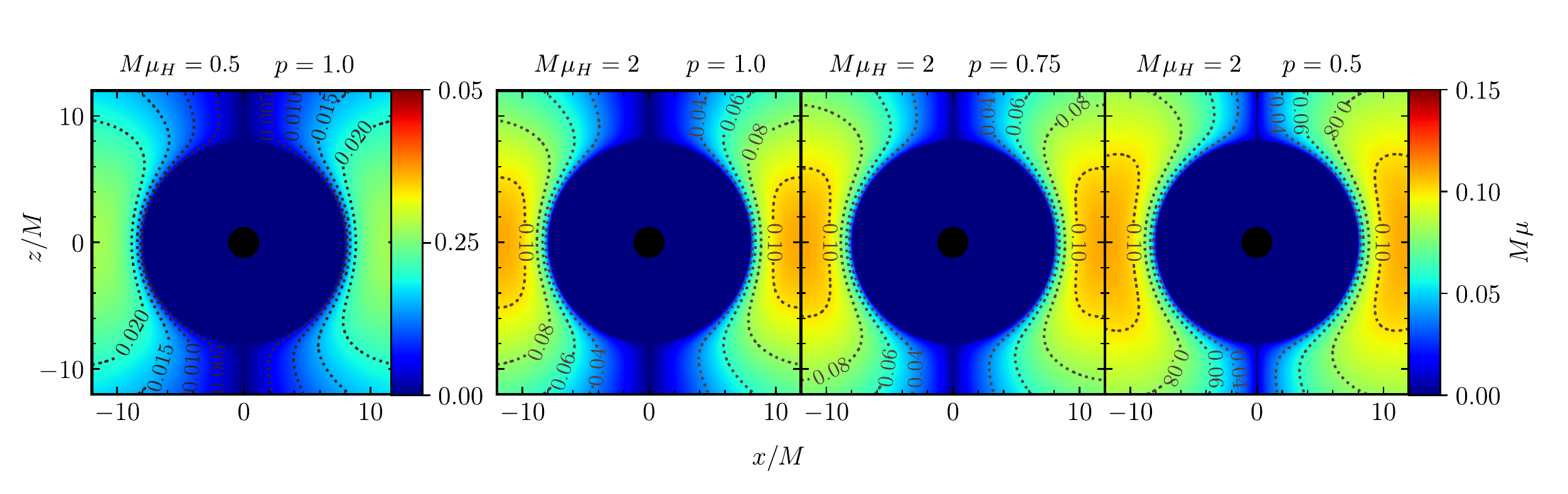}
  \caption{We illustrate the four configurations studied in the ``Shell-angular'' case, as listed in Table~\ref{table:growth}: top to bottom in the table maps from left to right in these plots. The smaller $p$ values towards the right correspond to a thicker plasma disk, which we find to be more efficient at supporting superradiant growth. Note the different color map in the left panel.}
  \label{fig:shell_angular}
\end{figure*}

\subsection{Mass profile configurations tested}
\label{sec:massprofile}
Eight of the Proca mass profiles that were tested in Ref.~\cite{Dima:2020rzg} are listed in Table~\ref{table:mass}, and shown in Fig.~\ref{fig:schema} (schematically) and Fig.~\ref{fig:mass_profile} (quantitatively). (Note that these are defined in Boyer-Lindquist coordinates, which are translated into the Kerr-Schild ones used in our simulations.) They are:

\begin{itemize}

\item[1)] Accretion: the radial mass profile
  \begin{equation}
    \mu_0(r) = \mu_H \left(\frac{r_{+}}{r}\right)^{\lambda/2}
    \label{eq:mu0}
  \end{equation}
  is meant to reproduce the qualitative features of spherically symmetric Bondi accretion~\cite{Bondi_accretion}. The parameters $\mu_H$ and $\lambda$ are a normalization mass and radial power-law index, respectively.

\item[2)] Accretion-angular: modifies model 1 by introducing a term proportional to $\sin \theta$, which changes the spherically symmetric mass profile into an axisymmetric disk centered at the equatorial plane. This mass potential mimics advection dominated accretion flows~\cite{ADAF_1994,ADAF_1995,ADAF_1995_2}. 

Model 2, when transformed to our Kerr-Schild coordinates, features a region immediately outside the BH horizon where the Proca mass sharply decreases to zero, which causes a discontinuity in the derivative of the Proca mass $\partial_a \left( \ln \mu \right)$ in \eqref{eqn-new_lorenz}, as illustrated in Fig.~\ref{fig:mu2deri}. Such a discontinuity triggers numerical instabilities in our simulations, so we remove it by adding a small constant mass $\mu_c = 0.02M^{-1}$ in our implementation of model 2. A smaller value of $\mu_c$ would cause the derivative to change too rapidly for our finite-difference evolution scheme in the problematic region. The chosen value of $\mu_c$ is well below the Proca mass $M\mu \sim 0.5$ corresponding to the strongest superradiance, so any superradiant growth related to this constant mass term will occur on timescales much longer than our simulations, and therefore it does not significantly affect our results (indeed in this case we do not observe superradiance, and would expect this change to only enhance it).

\item[3)] Accretion-constant: modifies model 1 by adding a further constant mass term $\mu_c$, which aims to capture the effect of the asympotic ISM density~\cite{Schnitzeler:2012jq}.

\item[4)] Accretion-angular-constant: has both the angular $|\sin\theta|$ dependence and a constant mass term $\mu_c$.

\item[5)] Shell: the radial mass profile
  \begin{equation}
    \mu_1(r) = \mu_H \sqrt{\Theta(r-r_0)\left(1-\frac{r_0}{r}\right)\left(\frac{r_0}{r}\right)^{\lambda}}
    \label{eq:mu1}
  \end{equation}
models the possibility that the accretion disk truncates at some radius near the BH. 
The truncation radius is set by the parameter $r_0$. 

In our numerical implementation, we replace the term $\Theta(r-r_0)(1-r_0/r)$ with a smooth sigmoid function ${[1+e^{-k(r-r'_0)/r'_0}]^{-1}}$. We choose $k$ and $r'_0$ to approximate the original profile as closely as possible, whilst still being numerically tractable with our finite difference scheme.

\item[6)] Shell-angular: modifies model 5 by introducing the angular dependence through a $|\sin \theta|^p$ term, similar to (but more general than) model 2.

\item[7)] Shell-constant: modifies model 5 by introducing a constant mass term $\mu_c$, as in model 3.
  
\item[8)] Shell-angular-constant: modifies model 5 by introducing both the angular $|\sin\theta|$ dependence and a constant mass term $\mu_c$, as in model 4.

\end{itemize}

The parameters used in each mass profile, and the exponential growth (or decay) rates found by fitting results from these simulations, are listed in Table~\ref{table:growth}.

\section{Results}
\label{sec:results}

In Fig.~\ref{fig:multipanel} we summarize the results of our simulations for the mass profiles listed in Tables~\ref{table:mass} and \ref{table:growth}. Each panel in Fig.~\ref{fig:multipanel} shows the Proca field energy as a function of time for a different mass profile.
All simulations use the same initial conditions and integration region as in the ``Uniform'' (constant Proca mass) case, which is shown as a dash-dotted gray line for comparison in all of the panels: see also Fig.~\ref{fig:constpic}, where we plot the Proca energy distribution at different time slices on the $x-y$ plane for the constant-mass case.\footnote{In the uniform case we choose $M\mu=0.5$, corresponding to the maximum possible Proca growth rate. In the cases where we add an asymptotic mass we choose $M\mu=0.42$ to match the value used in Dima et al.~\cite{Dima:2020rzg}, for which the uniform Proca rate would be slightly lower. The difference in the rates between the two values of $M\mu$ is negligible, and it would not noticeably change the ``reference'' dash-dotted gray lines in Fig.~\ref{fig:multipanel}.}

The two panels in the leftmost column correspond to the radial mass distributions $\mu_0(r)$ and $\mu_1(r)$ that approximate Bondi accretion and a spherical hollow shell, respectively. We then modify these distributions by adding an angular dependence (second column), a constant mass term (third column), or both (fourth column).

In simulations using Bondi accretion alone (either with or without an angular modification) the field energy decays in time.  However, by adding a constant mass term we observe a superradiant growth rate which is approximately half ($0.367\pm0.006$ and $0.539\pm0.010$ for the accretion-constant and accretion-angular-constant cases, respectively) of the constant-mass growth rate. These results suggest that a Bondi accretion profile alone is not enough to trigger superradiance, but that a nonzero asymptotic mass (provided e.g. by the ISM) may still trigger a superradiant growth of the Proca field, albeit with growth rates somewhat slower than the ideal constant-mass case.

We do find superradiance for the spherical shell distribution, which features a ``hollow'' inner region close to the BH where the Proca mass is suppressed. The boundary of this region at $r_0=8M$ provides a mirror-like structure that reflects the Proca field back to the ergoregion, helping to enhance the superradiant scattering of the field.  Having a sufficiently large radius for this region is found to be crucial: the dashed red line in the bottom-left panel shows that no superradiant growth is observed if the boundary is located at $r_0 = r_{\text{ISCO}} = 1.454M$.

When we add an angular modification of the spherical shell distribution, we find that the superradiant growth is strongly dependent on the thickness of the disk. The ``Shell-angular'' profiles we studied are shown in Fig.~\ref{fig:shell_angular}.

\begin{figure}[t]
    \includegraphics[width=0.48\textwidth]{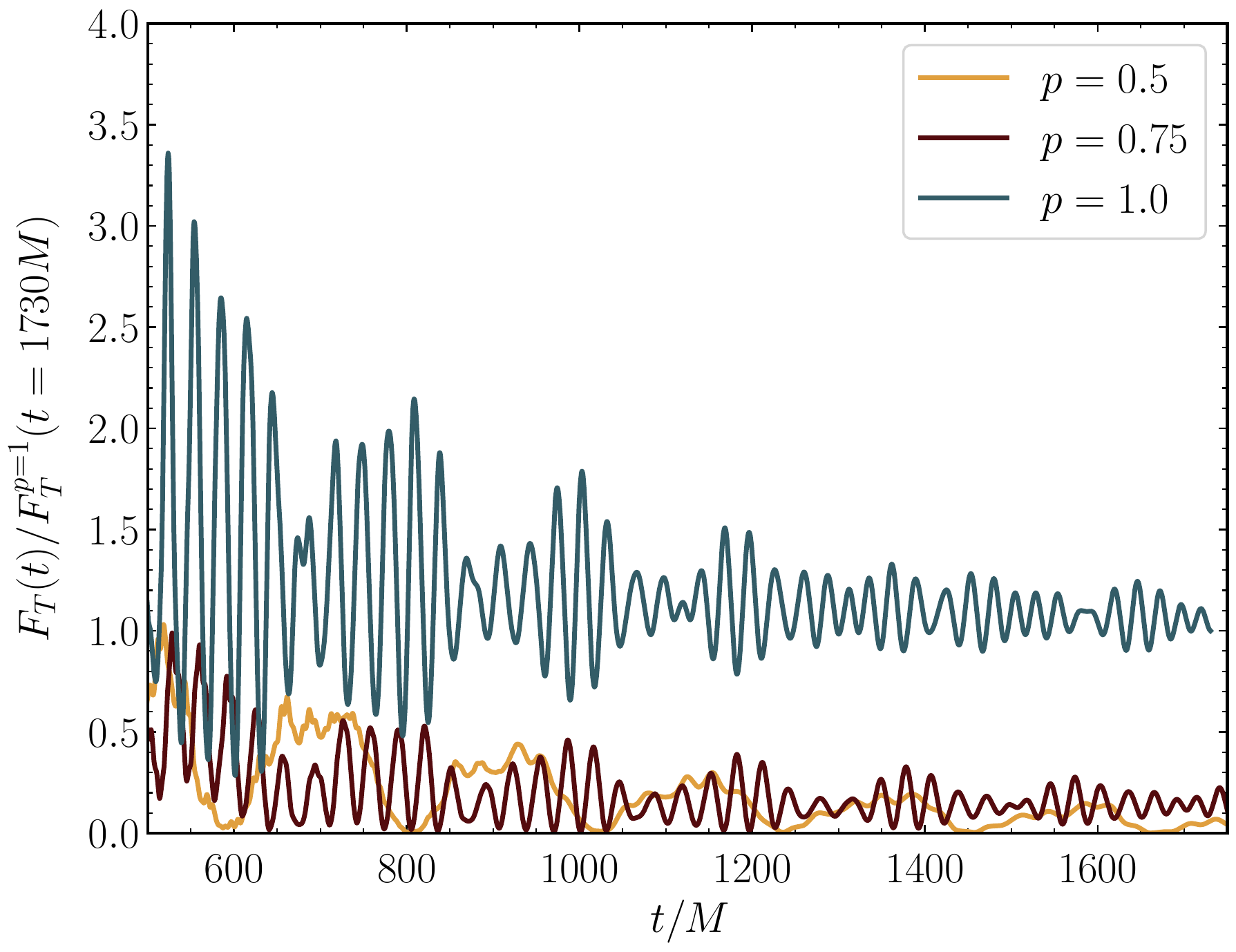}
    \caption{Energy flux out of a sphere at radius $r=144M$ for simulations with the ``Shell-angular'' profile, normalized to the final value of the flux in the $p=1$ case. After the field settles down from the initial evolution ($t>500M$), a net positive energy flux is still present, indicating ``leakage'' of the superradiant field due to the disk-like effective mass. The leakage, which is smaller for thicker disks (smaller values of $p$), can inhibit the growth of the condensate.}
    \label{fig:mu6flux}
\end{figure}

In thick disks, similar to those studied in Ref.~\cite{Dima:2020rzg}, we find that the superradiant instability can still be triggered. However, for even slightly thinner disks (achieved by adding a power $p$ to the $\sin{\theta}$ term, $\mu \sim |\sin{\theta}|^p$) the growth is lost and the mass of the superradiant Proca field decays exponentially. With this angular modification, the ``mirror'' structure disappears at the two poles on the $z$-axis, which allows leakage of the superradiant modes from the inner region.  The leakage is illustrated in Fig.~\ref{fig:mu6flux}, where we see a net positive outward energy flux in this configuration that increases for thinner disks. This implies that thinner disks would be less likely to support superradiant growth, as the Proca field is able to escape at the poles.

For the spherical shell distribution with an additional constant mass term (``Shell-constant'') we observe a superradiant growth rate $\sim 50\%$ greater than that of the constant mass case. For the shell distribution with both an angular dependence and a constant mass term (``Shell-angular-constant''), the superradiant growth rate is slightly slower, only $\sim 30\%$ greater than that of the constant mass case. Similar to the results we found in modifications of the Bondi accretion, these results suggest that a roughly constant mass term is key to restoring superradiant instabilities.  Of course, we should emphasize that we have in all cases chosen to add a constant mass term that is in an appropriate range to support superradiance on short timescales for the chosen BH mass, so this should be regarded as a ``best-case'' scenario. Where the mass is outside the optimal range, the timescale of superradiance will be highly suppressed, as it would be in the uniform mass case.

\section{Discussion} \label{maintext:discussion}
\label{sec:discussion}

In this work we have performed relativistic, nonlinear 3+1D evolutions of a massive vector field around a Kerr BH, with a spatially varying effective mass to mimic the case of photons interacting with a plasma.  We find that different models for the distribution of the plasma can both stall the superradiant build up, and actually enhance the superradiant growth rate. In particular, we find that for models 1 (Accretion) and 2 (Accretion-angular) no superradiant instability is observed; for models 3 (Accretion-constant) and 4 (Accretion-angular-constant) the superradiant growth is slower than in the constant Proca mass case; and for models 7 (Shell-constant) and 8 (Shell-angular-constant) the superradiant growth rate is enhanced. The outcome in the case of model 6 (Shell-angular) depends on the thickness of the disk, with thicker disks supporting a (slower than uniform) growth rate, and thinner disks leading to a decay of the superradiant mode due to leakage at the poles.

These results are broadly consistent with the previous work using scalar fields as a proxy~\cite{Dima:2020rzg}. In particular, they emphasize the importance of a nonzero asymptotic mass in confining the field sufficiently for the superradiant instability to take hold. Alternatively, the ``mirror-like'' structure provided by an inner cut-off in the plasma cloud can trigger and even provide enhanced rates of superradiance in comparison to the uniform mass case. In general, having merely a localized overdensity, such as an accretion spike or disk, is insufficient to obtain superradiant growth. On the other hand, the presence of such structures, in addition to a nonzero asymptotic mass or an inner cut-off, does not inhibit the growth. 

Our results confirm that vector and scalar fields have a similar response to nonuniform mass terms, and thus the analysis of ~\cite{Dima:2020rzg} indeed applies to the photon case by which their work was motivated. One aspect in which we were able to extend their studies was to investigate the impact of disk thickness on the superradiant growth rate. We confirm their finding that superradiance occurs for thick disks, but we have shown that thinner disks reduce the superradiant growth rate, and can lead to an overall decay of the bound Proca field. This is because they permit leakage of the field from the poles, which can dominate over the growth from superradiant scattering in the ergosphere.

Our Proca field evolution is performed on a fixed Kerr background, which is a very good approximation for the initial stages of superradiant growth, where there is a small density relative to the curvature scale of the BH ($M^2\rho \ll 1$ in geometrical units).  Performing the simulations on a dynamically evolving background would permit the study of the effects of backreaction of the field onto the BH at later stages, in particular the spin-down or spin-up of the BH from superradiance, or accretion of the superradiant Proca field onto the BH due to instabilities. Such simulations are feasible, although computationally more expensive. A cheaper alternative would be to simply adjust the mass $M$ and angular momentum $J$ of the fixed background Kerr metric at each timestep based on the measured flux of these quantities into the BH horizon.  This technique would be effective where the evolution is smooth and adiabatic, but not in the case of rapid and violent bursts.  We leave such studies of the late growth to future work.

To apply our results to the electromagnetic field inside a plasma one must assume that it is well-approximated by a simple massive Proca field, neglecting self-interactions and interactions with other fields.  This assumption would break down when the field becomes large enough to disrupt the plasma distribution, or to turn on higher-order interactions. The inclusion of a self-interaction term in the Proca field could approximate some of the additional effects of real electromagnetic fields inside plasmas, along with a time evolving (as well as spatially varying) value for the mass term which takes account of the superradiant growth.

\acknowledgments 
We acknowledge many helpful conversations and practical advice from Helvi Witek, who also generously provided us with her Mathematica notebook for the Proca evolution equations. 
Z.W., T.H. and E.B. are supported by NSF Grants No.~PHY-1912550, AST-2006538, PHY-090003 and PHY-20043, and NASA Grants No.~17-ATP17-0225, 19-ATP19-0051 and 20-LPS20-0011.
K.C. acknowledges funding from the European Research Council (ERC) under the European Unions Horizon 2020 research and innovation programme (grant agreement No 693024), and an STFC Ernest Rutherford Fellowship.
This research project was conducted using computational resources at the Maryland Advanced Research Computing Center (MARCC).  The authors acknowledge the Texas Advanced Computing Center (TACC) at The University of Texas at Austin for providing HPC resources that have contributed to the research results reported within this paper (URL: \url{http://www.tacc.utexas.edu})~\cite{10.1145/3311790.3396656}.
The project also used DiRAC resources under the projects ACSP218 and ACTP238 and PRACE resources under Grant Numbers 2020225359 and 2018194669. This DiRAC work was performed using the Cambridge Service for Data Driven Discovery (CSD3), part of which is operated by the University of Cambridge Research Computing on behalf of the STFC DiRAC HPC Facility (www.dirac.ac.uk). The DiRAC component of CSD3 was funded by BEIS capital funding via STFC capital grants ST/P002307/1 and ST/R002452/1 and STFC operations grant ST/R00689X/1. It also used the DiRAC at Durham facility managed by the Institute for Computational Cosmology on behalf of the STFC DiRAC HPC Facility (www.dirac.ac.uk). The equipment was funded by BEIS capital funding via STFC capital grants ST/P002293/1 and ST/R002371/1, Durham University and STFC operations grant ST/R000832/1. DiRAC is part of the National e-Infrastructure. The PRACE resources used were the GCS Supercomputer JUWELS at J\"ulich Supercomputing Centre(JCS) through the John von Neumann Institute for Computing (NIC), funded by the Gauss Centre for Supercomputing e.V. (\url{www.gauss-centre.eu}) and computer resources at SuperMUCNG, with technical support provided by the Leibniz Supercomputing Center.

\appendix

\section{Numerical relativity setup} \label{appendix:numerical}
\label{app:numerical}

\subsection{Numerical methods and code validation}

We use the {\sc GRChombo} numerical relativity framework~\cite{Clough:2015sqa,Andrade2021,Radia:2021smk}, but we compute the metric components and their derivatives analytically at each point rather than storing them on the grid. The evolution of the Proca field follows the standard method of lines, with Runge-Kutta time integration and fourth-order finite difference stencils.

As discussed in Sec.~\ref{sec:kerr}, we evolve the Proca field on the fixed background metric in Kerr-Schild coordinates.  
The metric is validated by checking that the numerically calculated Hamiltonian and momentum constraints converge to zero with increasing resolution, as do the time derivatives of the metric components, i.e. $\partial_t \gamma_{ij} = \partial_t K_{ij} = 0$ (calculated using the ADM expressions). This ensures that (ignoring the backreaction) the metric is indeed stationary in the chosen gauge, consistent with it being fixed over the field evolution.

\begin{figure}[t]
  \includegraphics[width=0.45\textwidth]{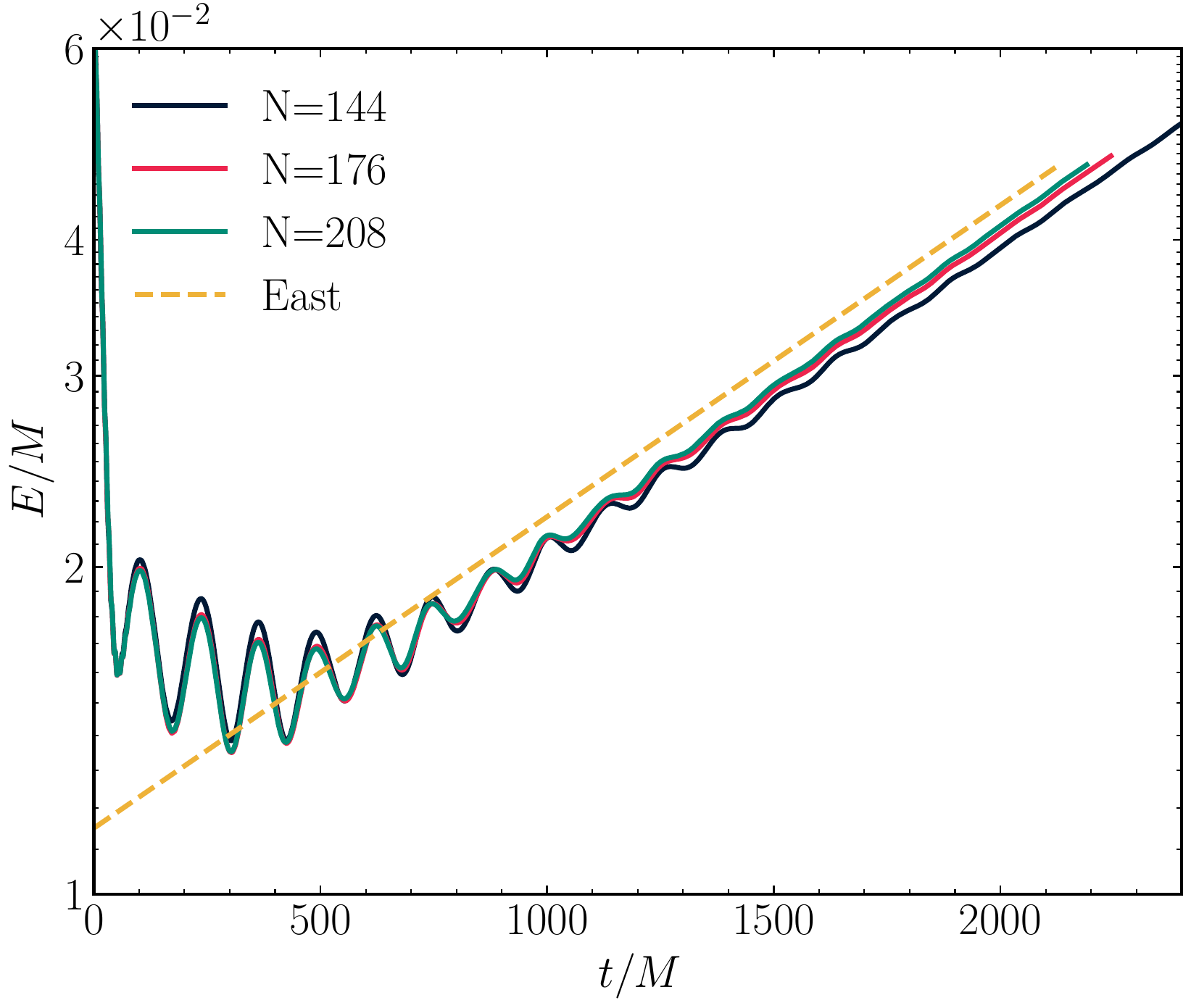}
    \caption{Convergence test of the constant Proca mass simulation.
    Blue, orange, and green curves are results obtained with $\Delta_1 =
    0.0208M$, $\Delta_2=0.0170M$, and $\Delta_3=0.0144M$ as coarsest resolutions,
    respectively. These
    resolutions correspond to 144, 176, and 208 boxes in the $x$- and $y$-direction
    (the grid in each direction extends over $384M$).
 The dashed orange line shows the superradiant growth rate obtained by East~\cite{East:2017mrj}.     }
    \label{fig:const_evolution}
\end{figure}

To verify the numerical scheme used in this paper, we first evolve the Proca field with a uniform Proca mass $M\mu=0.5$ using the initial conditions of \eqref{eqn:IC}. We then integrate $\rho$ in \eqref{eqn:energy} in the region between $r=2M$ and $r=144M$ to obtain the total matter energy of the Proca field. The resulting evolution is shown in Fig.~\ref{fig:const_evolution}.  At early times, i.e. before $t\sim500M$, nonsuperradiant modes in the initial data either fall into the BH or radiate away, causing the energy to decrease. After $t\sim500M$, the $m=1$ superradiant mode dominates the total energy growth. From the simulation at the highest resolution ($N=208$) we find a growth rate of $2 M \omega_{I} = (6.71\pm0.16)\times 10^{-4}$ for the energy, in good agreement with the value of $2 M \omega_{I} = 6.6 \times 10^{-4}$ found in Ref.~\cite{East:2017mrj}.

\begin{figure}
  \includegraphics[width=0.48\textwidth]{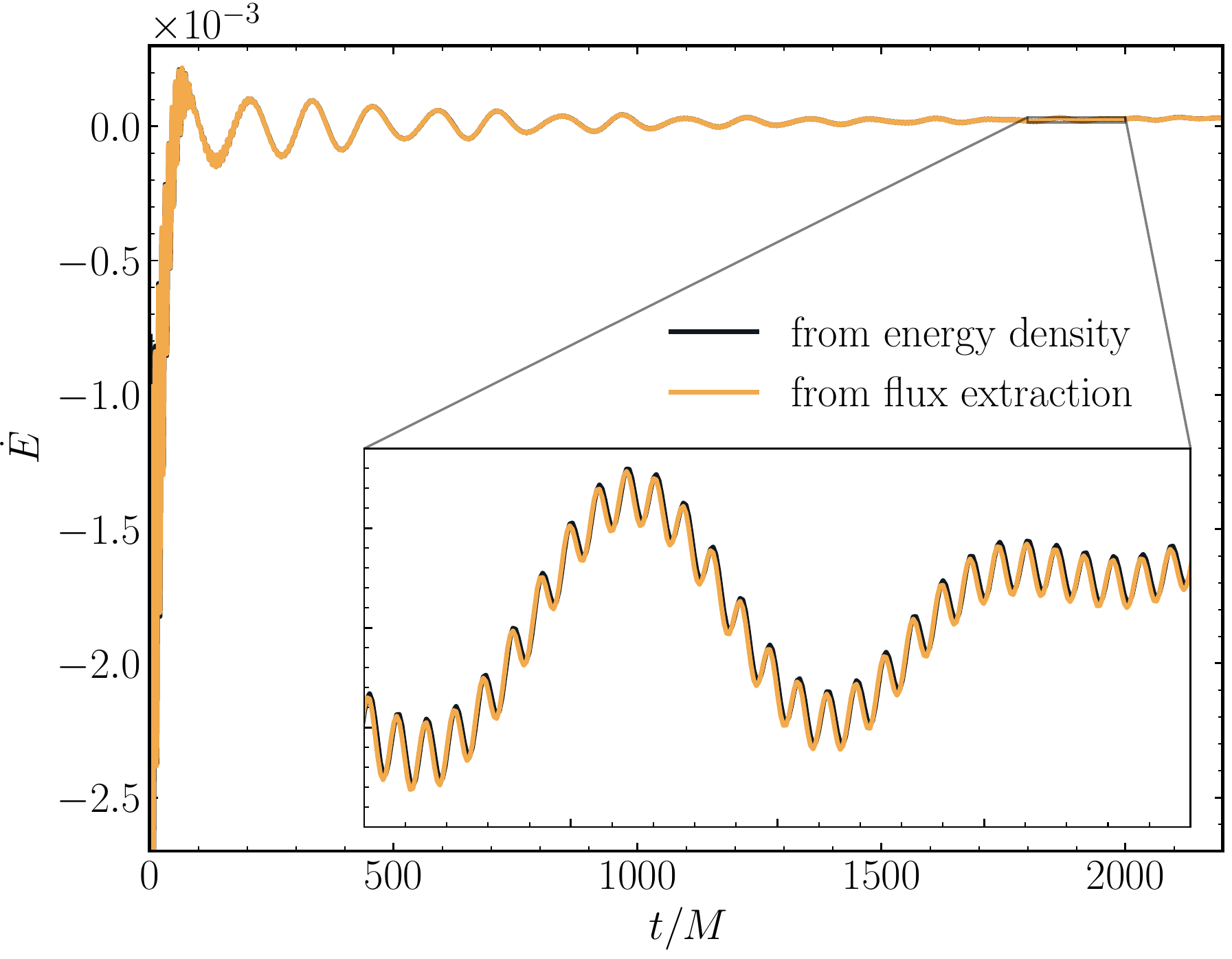}
    \caption{Consistency check between flux and volume integrals for a constant Proca mass simulation with intermediate resolution
    $\Delta_2=0.0170M$. In blue we show the time derivative of the integrated energy
    between $r=2M$ and $r=144M$; in orange, the ingoing flux at
    the two surfaces $r=2M$ and $r=144M$. The two curves lay on top of each
    other, consistent with energy conservation in the integration region.
    }
    \label{fig:flux}
\end{figure}

As the spacetime has a time-like Killing vector, the conservation equations (\ref{eqn:conservation}) must hold for all times. As a check of the quality of our simulations we can compare the right-hand and left-hand sides of the equality, thus checking that energy in the region between $r = 2 M$ and $r = 144M$ is conserved via the fluxes (see~\cite{Clough:2021qlv} for implementation details).  In Fig.~\ref{fig:flux} we demonstrate the good agreement between these two quantities.

\begin{figure}[ht]
  \includegraphics[width=0.48\textwidth]{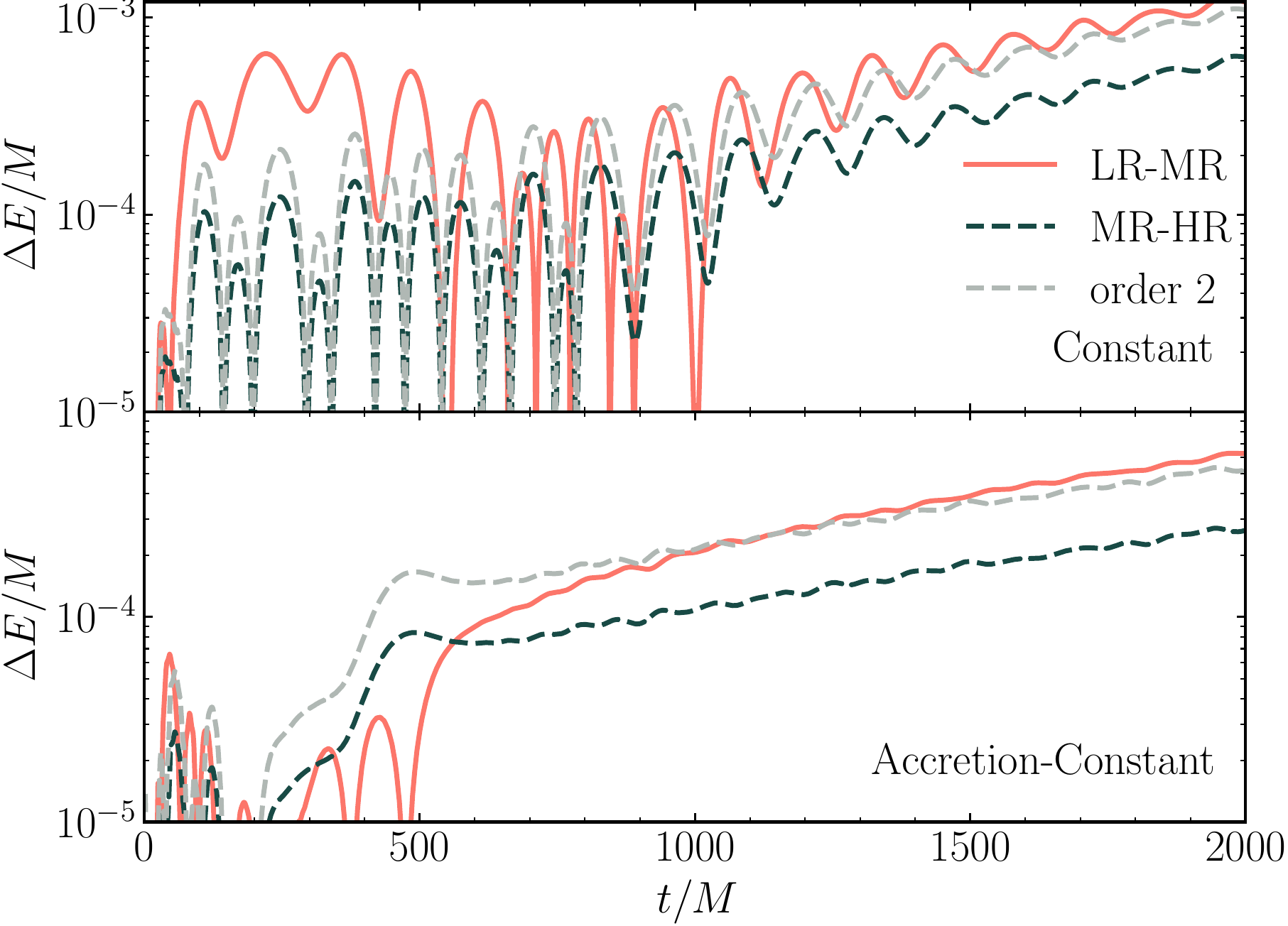}
    \caption{Convergence order for simulations with constant Proca mass (top panel) and for model 
    3 (bottom panel). In both cases, we show three runs with increasing resolution. The red lines show
    the energy difference between the low- and middle-resolution
    runs, and the dark dashed lines show the energy difference between the middle- and high-resolution
    runs. The dashed, light-gray curves show the difference between the
    middle- and high-resolution runs times the convergence factor $c(t)$, as
    calculated in \eqref{eqn:conv_factor} assuming second-order convergence, which is consistent with our results. }
    \label{fig:convergence}
\end{figure}

\subsection{Convergence testing}

We use a fixed hierarchy of grids with the largest box of size $L= 384M$, $N=144$ and 7 levels of refinement, with a resolution on the finest grid of
$\Delta x_{\rm fine} = 0.02083M$. We confirm that this resolution is sufficient by performing convergence tests as described in this section.

In Fig.~\ref{fig:const_evolution} we simulate the constant Proca mass scenario with different resolutions. There is some increase in the superradiant growth rate as we increase the resolution, but the change gets smaller as the resolution increases, and it is consistent with being in the convergent regime. Our results are also consistent with the growth rates found in previous work by East~\cite{East:2017mrj}.

We have performed convergence tests for both a constant Proca mass case and a representative, position-dependent mass profile (model 3), as shown in Fig.~\ref{fig:convergence}. For both cases we have computed the Proca field energy at three resolutions $\Delta_1$, $\Delta_2$ and $\Delta_3$.  For the constant Proca mass case (top panel) we used $\Delta_1 = 0.0208M$, $\Delta_2=0.0170M$, and $\Delta_3=0.0144M$, as shown in Fig.~\ref{fig:convergence} (top panel). For model 3 (bottom panel) we used $\Delta_1 = 0.0268M$, $\Delta_2 = 0.0208M$, and $\Delta_3=0.0144M$ (corresponding to $N=112$, $N=144$, and $N=176$).

The convergence factor is defined as the ratio of the relative differences between the solution at the low/medium and the middle/high resolutions:
\begin{equation}
    c(t) = \frac{||F_{\Delta_1} - F_{\Delta_2}||}{||F_{\Delta_2} -
    F_{\Delta_3}||}\,,
\end{equation}

In the limit $\Delta \to 0$ the convergence factor is expected to behave as
\begin{equation}
    \lim_{\Delta\to0}c(t) =
    \frac{\Delta_1^n-\Delta_2^n}{\Delta_2^n-\Delta_3^n}\,, \label{eqn:conv_factor}
\end{equation}
where $n$ is the order of the finite difference scheme used in the evolution. In Fig.~\ref{fig:convergence} we show that in the late-time superradiant phase of the evolution, the errors are consistent with convergence between second and third order. Whilst our finite difference stencils and time evolution are fourth order, interpolation at grid boundaries can reduce this to third order, and we also introduce errors from the ``lego sphere'' effect of zeroing cells inside the inner radius (meaning that the approximation of the volume over which the density is integrated does not converge at fourth order in resolution).

\subsection{Richardson extrapolation}
\label{sec:richardson}

In Table \ref{table:growth} of the main text, the error associated with the growth/decay rates of the Proca field energy was estimated using Richardson extrapolation. Given two sets of solutions with increasing resolutions $F_{\Delta_1}$ and $F_{\Delta_2}$, the error on the higher-resolution result can be estimated as
$\epsilon_{\Delta_2} \sim (F_{\Delta_1}-F_{\Delta_2})/(r^n - 1)$,
where $r=\Delta_1/\Delta_2$ is the ratio of the two resolutions, and $n$ is the convergence order of the evolution scheme used. To estimate the growth rate errors  in Table \ref{table:growth} we assume $n=4$, and we use data from two resolutions: $\Delta_1 = 2.67M$ and $\Delta_2 = 2.18M$ ($N=144$ and $N=176$).

\bibliography{mybib}

\begin{thebibliography}{78}%
\makeatletter
\providecommand \@ifxundefined [1]{%
 \@ifx{#1\undefined}
}%
\providecommand \@ifnum [1]{%
 \ifnum #1\expandafter \@firstoftwo
 \else \expandafter \@secondoftwo
 \fi
}%
\providecommand \@ifx [1]{%
 \ifx #1\expandafter \@firstoftwo
 \else \expandafter \@secondoftwo
 \fi
}%
\providecommand \natexlab [1]{#1}%
\providecommand \enquote  [1]{``#1''}%
\providecommand \bibnamefont  [1]{#1}%
\providecommand \bibfnamefont [1]{#1}%
\providecommand \citenamefont [1]{#1}%
\providecommand \href@noop [0]{\@secondoftwo}%
\providecommand \href [0]{\begingroup \@sanitize@url \@href}%
\providecommand \@href[1]{\@@startlink{#1}\@@href}%
\providecommand \@@href[1]{\endgroup#1\@@endlink}%
\providecommand \@sanitize@url [0]{\catcode `\\12\catcode `\$12\catcode
  `\&12\catcode `\#12\catcode `\^12\catcode `\_12\catcode `\%12\relax}%
\providecommand \@@startlink[1]{}%
\providecommand \@@endlink[0]{}%
\providecommand \url  [0]{\begingroup\@sanitize@url \@url }%
\providecommand \@url [1]{\endgroup\@href {#1}{\urlprefix }}%
\providecommand \urlprefix  [0]{URL }%
\providecommand \Eprint [0]{\href }%
\providecommand \doibase [0]{http://dx.doi.org/}%
\providecommand \selectlanguage [0]{\@gobble}%
\providecommand \bibinfo  [0]{\@secondoftwo}%
\providecommand \bibfield  [0]{\@secondoftwo}%
\providecommand \translation [1]{[#1]}%
\providecommand \BibitemOpen [0]{}%
\providecommand \bibitemStop [0]{}%
\providecommand \bibitemNoStop [0]{.\EOS\space}%
\providecommand \EOS [0]{\spacefactor3000\relax}%
\providecommand \BibitemShut  [1]{\csname bibitem#1\endcsname}%
\let\auto@bib@innerbib\@empty
%</preamble>
\bibitem [{\citenamefont {Penrose}\ and\ \citenamefont
  {Floyd}(1971)}]{Penrose:1971uk}%
  \BibitemOpen
  \bibfield  {author} {\bibinfo {author} {\bibfnamefont {R.}~\bibnamefont
  {Penrose}}\ and\ \bibinfo {author} {\bibfnamefont {R.~M.}\ \bibnamefont
  {Floyd}},\ }\href {\doibase 10.1038/physci229177a0} {\bibfield  {journal}
  {\bibinfo  {journal} {Nature}\ }\textbf {\bibinfo {volume} {229}},\ \bibinfo
  {pages} {177} (\bibinfo {year} {1971})}\BibitemShut {NoStop}%
\bibitem [{\citenamefont {Cardoso}\ \emph {et~al.}(2004)\citenamefont
  {Cardoso}, \citenamefont {Dias}, \citenamefont {Lemos},\ and\ \citenamefont
  {Yoshida}}]{Cardoso:2004nk}%
  \BibitemOpen
  \bibfield  {author} {\bibinfo {author} {\bibfnamefont {V.}~\bibnamefont
  {Cardoso}}, \bibinfo {author} {\bibfnamefont {O.~J.~C.}\ \bibnamefont
  {Dias}}, \bibinfo {author} {\bibfnamefont {J.~P.~S.}\ \bibnamefont {Lemos}},
  \ and\ \bibinfo {author} {\bibfnamefont {S.}~\bibnamefont {Yoshida}},\ }\href
  {\doibase 10.1103/PhysRevD.70.049903} {\bibfield  {journal} {\bibinfo
  {journal} {Phys. Rev. D}\ }\textbf {\bibinfo {volume} {70}},\ \bibinfo
  {pages} {044039} (\bibinfo {year} {2004})},\ \bibinfo {note} {[Erratum:
  Phys.Rev.D 70, 049903 (2004)]},\ \Eprint
  {http://arxiv.org/abs/hep-th/0404096} {arXiv:hep-th/0404096} \BibitemShut
  {NoStop}%
\bibitem [{\citenamefont {{Zel'Dovich}}(1971)}]{1971JETPL..14..180Z}%
  \BibitemOpen
  \bibfield  {author} {\bibinfo {author} {\bibfnamefont {Y.~B.}\ \bibnamefont
  {{Zel'Dovich}}},\ }\href@noop {} {\bibfield  {journal} {\bibinfo  {journal}
  {Soviet Journal of Experimental and Theoretical Physics Letters}\ }\textbf
  {\bibinfo {volume} {14}},\ \bibinfo {pages} {180} (\bibinfo {year}
  {1971})}\BibitemShut {NoStop}%
\bibitem [{\citenamefont {Witek}\ \emph {et~al.}(2013)\citenamefont {Witek},
  \citenamefont {Cardoso}, \citenamefont {Ishibashi},\ and\ \citenamefont
  {Sperhake}}]{Witek:2012tr}%
  \BibitemOpen
  \bibfield  {author} {\bibinfo {author} {\bibfnamefont {H.}~\bibnamefont
  {Witek}}, \bibinfo {author} {\bibfnamefont {V.}~\bibnamefont {Cardoso}},
  \bibinfo {author} {\bibfnamefont {A.}~\bibnamefont {Ishibashi}}, \ and\
  \bibinfo {author} {\bibfnamefont {U.}~\bibnamefont {Sperhake}},\ }\href
  {\doibase 10.1103/PhysRevD.87.043513} {\bibfield  {journal} {\bibinfo
  {journal} {Phys. Rev. D}\ }\textbf {\bibinfo {volume} {87}},\ \bibinfo
  {pages} {043513} (\bibinfo {year} {2013})},\ \Eprint
  {http://arxiv.org/abs/1212.0551} {arXiv:1212.0551 [gr-qc]} \BibitemShut
  {NoStop}%
\bibitem [{\citenamefont {East}\ \emph {et~al.}(2014)\citenamefont {East},
  \citenamefont {Ramazano\u{g}lu},\ and\ \citenamefont
  {Pretorius}}]{East:2013mfa}%
  \BibitemOpen
  \bibfield  {author} {\bibinfo {author} {\bibfnamefont {W.~E.}\ \bibnamefont
  {East}}, \bibinfo {author} {\bibfnamefont {F.~M.}\ \bibnamefont
  {Ramazano\u{g}lu}}, \ and\ \bibinfo {author} {\bibfnamefont {F.}~\bibnamefont
  {Pretorius}},\ }\href {\doibase 10.1103/PhysRevD.89.061503} {\bibfield
  {journal} {\bibinfo  {journal} {Phys. Rev. D}\ }\textbf {\bibinfo {volume}
  {89}},\ \bibinfo {pages} {061503} (\bibinfo {year} {2014})},\ \Eprint
  {http://arxiv.org/abs/1312.4529} {arXiv:1312.4529 [gr-qc]} \BibitemShut
  {NoStop}%
\bibitem [{\citenamefont {Okawa}\ \emph {et~al.}(2014)\citenamefont {Okawa},
  \citenamefont {Witek},\ and\ \citenamefont {Cardoso}}]{Okawa:2014nda}%
  \BibitemOpen
  \bibfield  {author} {\bibinfo {author} {\bibfnamefont {H.}~\bibnamefont
  {Okawa}}, \bibinfo {author} {\bibfnamefont {H.}~\bibnamefont {Witek}}, \ and\
  \bibinfo {author} {\bibfnamefont {V.}~\bibnamefont {Cardoso}},\ }\href
  {\doibase 10.1103/PhysRevD.89.104032} {\bibfield  {journal} {\bibinfo
  {journal} {Phys. Rev. D}\ }\textbf {\bibinfo {volume} {89}},\ \bibinfo
  {pages} {104032} (\bibinfo {year} {2014})},\ \Eprint
  {http://arxiv.org/abs/1401.1548} {arXiv:1401.1548 [gr-qc]} \BibitemShut
  {NoStop}%
\bibitem [{\citenamefont {Zilh\~ao}\ \emph {et~al.}(2015)\citenamefont
  {Zilh\~ao}, \citenamefont {Witek},\ and\ \citenamefont
  {Cardoso}}]{Zilhao:2015tya}%
  \BibitemOpen
  \bibfield  {author} {\bibinfo {author} {\bibfnamefont {M.}~\bibnamefont
  {Zilh\~ao}}, \bibinfo {author} {\bibfnamefont {H.}~\bibnamefont {Witek}}, \
  and\ \bibinfo {author} {\bibfnamefont {V.}~\bibnamefont {Cardoso}},\ }\href
  {\doibase 10.1088/0264-9381/32/23/234003} {\bibfield  {journal} {\bibinfo
  {journal} {Class. Quant. Grav.}\ }\textbf {\bibinfo {volume} {32}},\ \bibinfo
  {pages} {234003} (\bibinfo {year} {2015})},\ \Eprint
  {http://arxiv.org/abs/1505.00797} {arXiv:1505.00797 [gr-qc]} \BibitemShut
  {NoStop}%
\bibitem [{\citenamefont {East}\ and\ \citenamefont
  {Pretorius}(2017)}]{East:2017ovw}%
  \BibitemOpen
  \bibfield  {author} {\bibinfo {author} {\bibfnamefont {W.~E.}\ \bibnamefont
  {East}}\ and\ \bibinfo {author} {\bibfnamefont {F.}~\bibnamefont
  {Pretorius}},\ }\href {\doibase 10.1103/PhysRevLett.119.041101} {\bibfield
  {journal} {\bibinfo  {journal} {Phys. Rev. Lett.}\ }\textbf {\bibinfo
  {volume} {119}},\ \bibinfo {pages} {041101} (\bibinfo {year} {2017})},\
  \Eprint {http://arxiv.org/abs/1704.04791} {arXiv:1704.04791 [gr-qc]}
  \BibitemShut {NoStop}%
\bibitem [{\citenamefont {East}(2017)}]{East:2017mrj}%
  \BibitemOpen
  \bibfield  {author} {\bibinfo {author} {\bibfnamefont {W.~E.}\ \bibnamefont
  {East}},\ }\href {\doibase 10.1103/PhysRevD.96.024004} {\bibfield  {journal}
  {\bibinfo  {journal} {Phys. Rev. D}\ }\textbf {\bibinfo {volume} {96}},\
  \bibinfo {pages} {024004} (\bibinfo {year} {2017})},\ \Eprint
  {http://arxiv.org/abs/1705.01544} {arXiv:1705.01544 [gr-qc]} \BibitemShut
  {NoStop}%
\bibitem [{\citenamefont {East}(2018)}]{East:2018glu}%
  \BibitemOpen
  \bibfield  {author} {\bibinfo {author} {\bibfnamefont {W.~E.}\ \bibnamefont
  {East}},\ }\href {\doibase 10.1103/PhysRevLett.121.131104} {\bibfield
  {journal} {\bibinfo  {journal} {Phys. Rev. Lett.}\ }\textbf {\bibinfo
  {volume} {121}},\ \bibinfo {pages} {131104} (\bibinfo {year} {2018})},\
  \Eprint {http://arxiv.org/abs/1807.00043} {arXiv:1807.00043 [gr-qc]}
  \BibitemShut {NoStop}%
\bibitem [{\citenamefont {Press}\ and\ \citenamefont
  {Teukolsky}(1972)}]{Press:1972zz}%
  \BibitemOpen
  \bibfield  {author} {\bibinfo {author} {\bibfnamefont {W.~H.}\ \bibnamefont
  {Press}}\ and\ \bibinfo {author} {\bibfnamefont {S.~A.}\ \bibnamefont
  {Teukolsky}},\ }\href {\doibase 10.1038/238211a0} {\bibfield  {journal}
  {\bibinfo  {journal} {Nature}\ }\textbf {\bibinfo {volume} {238}},\ \bibinfo
  {pages} {211} (\bibinfo {year} {1972})}\BibitemShut {NoStop}%
\bibitem [{\citenamefont {Deruelle}\ and\ \citenamefont
  {Ruffini}(1974)}]{Deruelle:1974zy}%
  \BibitemOpen
  \bibfield  {author} {\bibinfo {author} {\bibfnamefont {N.}~\bibnamefont
  {Deruelle}}\ and\ \bibinfo {author} {\bibfnamefont {R.}~\bibnamefont
  {Ruffini}},\ }\href {\doibase 10.1016/0370-2693(74)90119-1} {\bibfield
  {journal} {\bibinfo  {journal} {Phys. Lett. B}\ }\textbf {\bibinfo {volume}
  {52}},\ \bibinfo {pages} {437} (\bibinfo {year} {1974})}\BibitemShut
  {NoStop}%
\bibitem [{\citenamefont {Damour}\ \emph {et~al.}(1976)\citenamefont {Damour},
  \citenamefont {Deruelle},\ and\ \citenamefont {Ruffini}}]{Damour:1976kh}%
  \BibitemOpen
  \bibfield  {author} {\bibinfo {author} {\bibfnamefont {T.}~\bibnamefont
  {Damour}}, \bibinfo {author} {\bibfnamefont {N.}~\bibnamefont {Deruelle}}, \
  and\ \bibinfo {author} {\bibfnamefont {R.}~\bibnamefont {Ruffini}},\ }\href
  {\doibase 10.1007/BF02725534} {\bibfield  {journal} {\bibinfo  {journal}
  {Lett. Nuovo Cim.}\ }\textbf {\bibinfo {volume} {15}},\ \bibinfo {pages}
  {257} (\bibinfo {year} {1976})}\BibitemShut {NoStop}%
\bibitem [{\citenamefont {Detweiler}(1980)}]{Detweiler:1980uk}%
  \BibitemOpen
  \bibfield  {author} {\bibinfo {author} {\bibfnamefont {S.~L.}\ \bibnamefont
  {Detweiler}},\ }\href {\doibase 10.1103/PhysRevD.22.2323} {\bibfield
  {journal} {\bibinfo  {journal} {Phys. Rev. D}\ }\textbf {\bibinfo {volume}
  {22}},\ \bibinfo {pages} {2323} (\bibinfo {year} {1980})}\BibitemShut
  {NoStop}%
\bibitem [{\citenamefont {Gaina}\ and\ \citenamefont
  {Zaslavsky}(1992)}]{Gaina:1992nx}%
  \BibitemOpen
  \bibfield  {author} {\bibinfo {author} {\bibfnamefont {A.~B.}\ \bibnamefont
  {Gaina}}\ and\ \bibinfo {author} {\bibfnamefont {O.}~\bibnamefont
  {Zaslavsky}},\ }\href {\doibase 10.1088/0264-9381/9/3/009} {\bibfield
  {journal} {\bibinfo  {journal} {Class. Quant. Grav.}\ }\textbf {\bibinfo
  {volume} {9}},\ \bibinfo {pages} {667} (\bibinfo {year} {1992})}\BibitemShut
  {NoStop}%
\bibitem [{\citenamefont {Zouros}\ and\ \citenamefont
  {Eardley}(1979)}]{Zouros:1979iw}%
  \BibitemOpen
  \bibfield  {author} {\bibinfo {author} {\bibfnamefont {T.}~\bibnamefont
  {Zouros}}\ and\ \bibinfo {author} {\bibfnamefont {D.}~\bibnamefont
  {Eardley}},\ }\href {\doibase 10.1016/0003-4916(79)90237-9} {\bibfield
  {journal} {\bibinfo  {journal} {Annals Phys.}\ }\textbf {\bibinfo {volume}
  {118}},\ \bibinfo {pages} {139} (\bibinfo {year} {1979})}\BibitemShut
  {NoStop}%
\bibitem [{\citenamefont {Lasenby}\ \emph {et~al.}(2005)\citenamefont
  {Lasenby}, \citenamefont {Doran}, \citenamefont {Pritchard}, \citenamefont
  {Caceres},\ and\ \citenamefont {Dolan}}]{Lasenby:2002mc}%
  \BibitemOpen
  \bibfield  {author} {\bibinfo {author} {\bibfnamefont {A.}~\bibnamefont
  {Lasenby}}, \bibinfo {author} {\bibfnamefont {C.}~\bibnamefont {Doran}},
  \bibinfo {author} {\bibfnamefont {J.}~\bibnamefont {Pritchard}}, \bibinfo
  {author} {\bibfnamefont {A.}~\bibnamefont {Caceres}}, \ and\ \bibinfo
  {author} {\bibfnamefont {S.}~\bibnamefont {Dolan}},\ }\href {\doibase
  10.1103/PhysRevD.72.105014} {\bibfield  {journal} {\bibinfo  {journal} {Phys.
  Rev. D}\ }\textbf {\bibinfo {volume} {72}},\ \bibinfo {pages} {105014}
  (\bibinfo {year} {2005})},\ \Eprint {http://arxiv.org/abs/gr-qc/0209090}
  {arXiv:gr-qc/0209090} \BibitemShut {NoStop}%
\bibitem [{\citenamefont {Cardoso}\ and\ \citenamefont
  {Yoshida}(2005)}]{Cardoso:2005vk}%
  \BibitemOpen
  \bibfield  {author} {\bibinfo {author} {\bibfnamefont {V.}~\bibnamefont
  {Cardoso}}\ and\ \bibinfo {author} {\bibfnamefont {S.}~\bibnamefont
  {Yoshida}},\ }\href {\doibase 10.1088/1126-6708/2005/07/009} {\bibfield
  {journal} {\bibinfo  {journal} {JHEP}\ }\textbf {\bibinfo {volume} {07}},\
  \bibinfo {pages} {009} (\bibinfo {year} {2005})},\ \Eprint
  {http://arxiv.org/abs/hep-th/0502206} {arXiv:hep-th/0502206} \BibitemShut
  {NoStop}%
\bibitem [{\citenamefont {Dolan}(2007)}]{Dolan:2007mj}%
  \BibitemOpen
  \bibfield  {author} {\bibinfo {author} {\bibfnamefont {S.~R.}\ \bibnamefont
  {Dolan}},\ }\href {\doibase 10.1103/PhysRevD.76.084001} {\bibfield  {journal}
  {\bibinfo  {journal} {Phys. Rev. D}\ }\textbf {\bibinfo {volume} {76}},\
  \bibinfo {pages} {084001} (\bibinfo {year} {2007})},\ \Eprint
  {http://arxiv.org/abs/0705.2880} {arXiv:0705.2880 [gr-qc]} \BibitemShut
  {NoStop}%
\bibitem [{\citenamefont {Grain}\ and\ \citenamefont
  {Barrau}(2008)}]{Grain:2007gn}%
  \BibitemOpen
  \bibfield  {author} {\bibinfo {author} {\bibfnamefont {J.}~\bibnamefont
  {Grain}}\ and\ \bibinfo {author} {\bibfnamefont {A.}~\bibnamefont {Barrau}},\
  }\href {\doibase 10.1140/epjc/s10052-007-0494-1} {\bibfield  {journal}
  {\bibinfo  {journal} {Eur. Phys. J. C}\ }\textbf {\bibinfo {volume} {53}},\
  \bibinfo {pages} {641} (\bibinfo {year} {2008})},\ \Eprint
  {http://arxiv.org/abs/hep-th/0701265} {arXiv:hep-th/0701265} \BibitemShut
  {NoStop}%
\bibitem [{\citenamefont {Arvanitaki}\ \emph {et~al.}(2010)\citenamefont
  {Arvanitaki}, \citenamefont {Dimopoulos}, \citenamefont {Dubovsky},
  \citenamefont {Kaloper},\ and\ \citenamefont
  {March-Russell}}]{Arvanitaki:2009fg}%
  \BibitemOpen
  \bibfield  {author} {\bibinfo {author} {\bibfnamefont {A.}~\bibnamefont
  {Arvanitaki}}, \bibinfo {author} {\bibfnamefont {S.}~\bibnamefont
  {Dimopoulos}}, \bibinfo {author} {\bibfnamefont {S.}~\bibnamefont
  {Dubovsky}}, \bibinfo {author} {\bibfnamefont {N.}~\bibnamefont {Kaloper}}, \
  and\ \bibinfo {author} {\bibfnamefont {J.}~\bibnamefont {March-Russell}},\
  }\href {\doibase 10.1103/PhysRevD.81.123530} {\bibfield  {journal} {\bibinfo
  {journal} {Phys. Rev. D}\ }\textbf {\bibinfo {volume} {81}},\ \bibinfo
  {pages} {123530} (\bibinfo {year} {2010})},\ \Eprint
  {http://arxiv.org/abs/0905.4720} {arXiv:0905.4720 [hep-th]} \BibitemShut
  {NoStop}%
\bibitem [{\citenamefont {Arvanitaki}\ and\ \citenamefont
  {Dubovsky}(2011)}]{Arvanitaki:2010sy}%
  \BibitemOpen
  \bibfield  {author} {\bibinfo {author} {\bibfnamefont {A.}~\bibnamefont
  {Arvanitaki}}\ and\ \bibinfo {author} {\bibfnamefont {S.}~\bibnamefont
  {Dubovsky}},\ }\href {\doibase 10.1103/PhysRevD.83.044026} {\bibfield
  {journal} {\bibinfo  {journal} {Phys. Rev. D}\ }\textbf {\bibinfo {volume}
  {83}},\ \bibinfo {pages} {044026} (\bibinfo {year} {2011})},\ \Eprint
  {http://arxiv.org/abs/1004.3558} {arXiv:1004.3558 [hep-th]} \BibitemShut
  {NoStop}%
\bibitem [{\citenamefont {Dolan}(2013{\natexlab{a}})}]{Dolan:2012yt}%
  \BibitemOpen
  \bibfield  {author} {\bibinfo {author} {\bibfnamefont {S.~R.}\ \bibnamefont
  {Dolan}},\ }\href {\doibase 10.1103/PhysRevD.87.124026} {\bibfield  {journal}
  {\bibinfo  {journal} {Phys. Rev. D}\ }\textbf {\bibinfo {volume} {87}},\
  \bibinfo {pages} {124026} (\bibinfo {year} {2013}{\natexlab{a}})},\ \Eprint
  {http://arxiv.org/abs/1212.1477} {arXiv:1212.1477 [gr-qc]} \BibitemShut
  {NoStop}%
\bibitem [{\citenamefont {Yoshino}\ and\ \citenamefont
  {Kodama}(2014)}]{Yoshino:2013ofa}%
  \BibitemOpen
  \bibfield  {author} {\bibinfo {author} {\bibfnamefont {H.}~\bibnamefont
  {Yoshino}}\ and\ \bibinfo {author} {\bibfnamefont {H.}~\bibnamefont
  {Kodama}},\ }\href {\doibase 10.1093/ptep/ptu029} {\bibfield  {journal}
  {\bibinfo  {journal} {PTEP}\ }\textbf {\bibinfo {volume} {2014}},\ \bibinfo
  {pages} {043E02} (\bibinfo {year} {2014})},\ \Eprint
  {http://arxiv.org/abs/1312.2326} {arXiv:1312.2326 [gr-qc]} \BibitemShut
  {NoStop}%
\bibitem [{\citenamefont {Brito}\ \emph
  {et~al.}(2015{\natexlab{a}})\citenamefont {Brito}, \citenamefont {Cardoso},\
  and\ \citenamefont {Pani}}]{Brito:2014wla}%
  \BibitemOpen
  \bibfield  {author} {\bibinfo {author} {\bibfnamefont {R.}~\bibnamefont
  {Brito}}, \bibinfo {author} {\bibfnamefont {V.}~\bibnamefont {Cardoso}}, \
  and\ \bibinfo {author} {\bibfnamefont {P.}~\bibnamefont {Pani}},\ }\href
  {\doibase 10.1088/0264-9381/32/13/134001} {\bibfield  {journal} {\bibinfo
  {journal} {Class. Quant. Grav.}\ }\textbf {\bibinfo {volume} {32}},\ \bibinfo
  {pages} {134001} (\bibinfo {year} {2015}{\natexlab{a}})},\ \Eprint
  {http://arxiv.org/abs/1411.0686} {arXiv:1411.0686 [gr-qc]} \BibitemShut
  {NoStop}%
\bibitem [{\citenamefont {Arvanitaki}\ \emph {et~al.}(2015)\citenamefont
  {Arvanitaki}, \citenamefont {Baryakhtar},\ and\ \citenamefont
  {Huang}}]{Arvanitaki:2014wva}%
  \BibitemOpen
  \bibfield  {author} {\bibinfo {author} {\bibfnamefont {A.}~\bibnamefont
  {Arvanitaki}}, \bibinfo {author} {\bibfnamefont {M.}~\bibnamefont
  {Baryakhtar}}, \ and\ \bibinfo {author} {\bibfnamefont {X.}~\bibnamefont
  {Huang}},\ }\href {\doibase 10.1103/PhysRevD.91.084011} {\bibfield  {journal}
  {\bibinfo  {journal} {Phys. Rev. D}\ }\textbf {\bibinfo {volume} {91}},\
  \bibinfo {pages} {084011} (\bibinfo {year} {2015})},\ \Eprint
  {http://arxiv.org/abs/1411.2263} {arXiv:1411.2263 [hep-ph]} \BibitemShut
  {NoStop}%
\bibitem [{\citenamefont {Arvanitaki}\ \emph {et~al.}(2017)\citenamefont
  {Arvanitaki}, \citenamefont {Baryakhtar}, \citenamefont {Dimopoulos},
  \citenamefont {Dubovsky},\ and\ \citenamefont
  {Lasenby}}]{Arvanitaki:2016qwi}%
  \BibitemOpen
  \bibfield  {author} {\bibinfo {author} {\bibfnamefont {A.}~\bibnamefont
  {Arvanitaki}}, \bibinfo {author} {\bibfnamefont {M.}~\bibnamefont
  {Baryakhtar}}, \bibinfo {author} {\bibfnamefont {S.}~\bibnamefont
  {Dimopoulos}}, \bibinfo {author} {\bibfnamefont {S.}~\bibnamefont
  {Dubovsky}}, \ and\ \bibinfo {author} {\bibfnamefont {R.}~\bibnamefont
  {Lasenby}},\ }\href {\doibase 10.1103/PhysRevD.95.043001} {\bibfield
  {journal} {\bibinfo  {journal} {Phys. Rev. D}\ }\textbf {\bibinfo {volume}
  {95}},\ \bibinfo {pages} {043001} (\bibinfo {year} {2017})},\ \Eprint
  {http://arxiv.org/abs/1604.03958} {arXiv:1604.03958 [hep-ph]} \BibitemShut
  {NoStop}%
\bibitem [{\citenamefont {Cardoso}\ \emph {et~al.}(2018)\citenamefont
  {Cardoso}, \citenamefont {Dias}, \citenamefont {Hartnett}, \citenamefont
  {Middleton}, \citenamefont {Pani},\ and\ \citenamefont
  {Santos}}]{Cardoso:2018tly}%
  \BibitemOpen
  \bibfield  {author} {\bibinfo {author} {\bibfnamefont {V.}~\bibnamefont
  {Cardoso}}, \bibinfo {author} {\bibfnamefont {O.~J.~C.}\ \bibnamefont
  {Dias}}, \bibinfo {author} {\bibfnamefont {G.~S.}\ \bibnamefont {Hartnett}},
  \bibinfo {author} {\bibfnamefont {M.}~\bibnamefont {Middleton}}, \bibinfo
  {author} {\bibfnamefont {P.}~\bibnamefont {Pani}}, \ and\ \bibinfo {author}
  {\bibfnamefont {J.~E.}\ \bibnamefont {Santos}},\ }\href {\doibase
  10.1088/1475-7516/2018/03/043} {\bibfield  {journal} {\bibinfo  {journal}
  {JCAP}\ }\textbf {\bibinfo {volume} {03}},\ \bibinfo {pages} {043} (\bibinfo
  {year} {2018})},\ \Eprint {http://arxiv.org/abs/1801.01420} {arXiv:1801.01420
  [gr-qc]} \BibitemShut {NoStop}%
\bibitem [{\citenamefont {Frolov}\ \emph {et~al.}(2018)\citenamefont {Frolov},
  \citenamefont {Krtou\v{s}}, \citenamefont {Kubiz\v{n}\'ak},\ and\
  \citenamefont {Santos}}]{Frolov:2018ezx}%
  \BibitemOpen
  \bibfield  {author} {\bibinfo {author} {\bibfnamefont {V.~P.}\ \bibnamefont
  {Frolov}}, \bibinfo {author} {\bibfnamefont {P.}~\bibnamefont {Krtou\v{s}}},
  \bibinfo {author} {\bibfnamefont {D.}~\bibnamefont {Kubiz\v{n}\'ak}}, \ and\
  \bibinfo {author} {\bibfnamefont {J.~E.}\ \bibnamefont {Santos}},\ }\href
  {\doibase 10.1103/PhysRevLett.120.231103} {\bibfield  {journal} {\bibinfo
  {journal} {Phys. Rev. Lett.}\ }\textbf {\bibinfo {volume} {120}},\ \bibinfo
  {pages} {231103} (\bibinfo {year} {2018})},\ \Eprint
  {http://arxiv.org/abs/1804.00030} {arXiv:1804.00030 [hep-th]} \BibitemShut
  {NoStop}%
\bibitem [{\citenamefont {Dolan}(2018)}]{Dolan:2018dqv}%
  \BibitemOpen
  \bibfield  {author} {\bibinfo {author} {\bibfnamefont {S.~R.}\ \bibnamefont
  {Dolan}},\ }\href {\doibase 10.1103/PhysRevD.98.104006} {\bibfield  {journal}
  {\bibinfo  {journal} {Phys. Rev. D}\ }\textbf {\bibinfo {volume} {98}},\
  \bibinfo {pages} {104006} (\bibinfo {year} {2018})},\ \Eprint
  {http://arxiv.org/abs/1806.01604} {arXiv:1806.01604 [gr-qc]} \BibitemShut
  {NoStop}%
\bibitem [{\citenamefont {Ficarra}\ \emph {et~al.}(2019)\citenamefont
  {Ficarra}, \citenamefont {Pani},\ and\ \citenamefont
  {Witek}}]{Ficarra:2018rfu}%
  \BibitemOpen
  \bibfield  {author} {\bibinfo {author} {\bibfnamefont {G.}~\bibnamefont
  {Ficarra}}, \bibinfo {author} {\bibfnamefont {P.}~\bibnamefont {Pani}}, \
  and\ \bibinfo {author} {\bibfnamefont {H.}~\bibnamefont {Witek}},\ }\href
  {\doibase 10.1103/PhysRevD.99.104019} {\bibfield  {journal} {\bibinfo
  {journal} {Phys. Rev. D}\ }\textbf {\bibinfo {volume} {99}},\ \bibinfo
  {pages} {104019} (\bibinfo {year} {2019})},\ \Eprint
  {http://arxiv.org/abs/1812.02758} {arXiv:1812.02758 [gr-qc]} \BibitemShut
  {NoStop}%
\bibitem [{\citenamefont {Siemonsen}\ and\ \citenamefont
  {East}(2020)}]{Siemonsen:2019ebd}%
  \BibitemOpen
  \bibfield  {author} {\bibinfo {author} {\bibfnamefont {N.}~\bibnamefont
  {Siemonsen}}\ and\ \bibinfo {author} {\bibfnamefont {W.~E.}\ \bibnamefont
  {East}},\ }\href {\doibase 10.1103/PhysRevD.101.024019} {\bibfield  {journal}
  {\bibinfo  {journal} {Phys. Rev. D}\ }\textbf {\bibinfo {volume} {101}},\
  \bibinfo {pages} {024019} (\bibinfo {year} {2020})},\ \Eprint
  {http://arxiv.org/abs/1910.09476} {arXiv:1910.09476 [gr-qc]} \BibitemShut
  {NoStop}%
\bibitem [{\citenamefont {Creci}\ \emph {et~al.}(2020)\citenamefont {Creci},
  \citenamefont {Vandoren},\ and\ \citenamefont {Witek}}]{Creci:2020mfg}%
  \BibitemOpen
  \bibfield  {author} {\bibinfo {author} {\bibfnamefont {G.}~\bibnamefont
  {Creci}}, \bibinfo {author} {\bibfnamefont {S.}~\bibnamefont {Vandoren}}, \
  and\ \bibinfo {author} {\bibfnamefont {H.}~\bibnamefont {Witek}},\ }\href
  {\doibase 10.1103/PhysRevD.101.124051} {\bibfield  {journal} {\bibinfo
  {journal} {Phys. Rev. D}\ }\textbf {\bibinfo {volume} {101}},\ \bibinfo
  {pages} {124051} (\bibinfo {year} {2020})},\ \Eprint
  {http://arxiv.org/abs/2004.05178} {arXiv:2004.05178 [gr-qc]} \BibitemShut
  {NoStop}%
\bibitem [{\citenamefont {Brito}\ \emph
  {et~al.}(2015{\natexlab{b}})\citenamefont {Brito}, \citenamefont {Cardoso},\
  and\ \citenamefont {Pani}}]{Brito:2015oca}%
  \BibitemOpen
  \bibfield  {author} {\bibinfo {author} {\bibfnamefont {R.}~\bibnamefont
  {Brito}}, \bibinfo {author} {\bibfnamefont {V.}~\bibnamefont {Cardoso}}, \
  and\ \bibinfo {author} {\bibfnamefont {P.}~\bibnamefont {Pani}},\ }\href
  {\doibase 10.1007/978-3-319-19000-6} {\bibfield  {journal} {\bibinfo
  {journal} {Lect. Notes Phys.}\ }\textbf {\bibinfo {volume} {906}},\ \bibinfo
  {pages} {pp.1} (\bibinfo {year} {2015}{\natexlab{b}})},\ \Eprint
  {http://arxiv.org/abs/1501.06570} {arXiv:1501.06570 [gr-qc]} \BibitemShut
  {NoStop}%
\bibitem [{\citenamefont {Baryakhtar}\ \emph {et~al.}(2021)\citenamefont
  {Baryakhtar}, \citenamefont {Galanis}, \citenamefont {Lasenby},\ and\
  \citenamefont {Simon}}]{Baryakhtar:2020gao}%
  \BibitemOpen
  \bibfield  {author} {\bibinfo {author} {\bibfnamefont {M.}~\bibnamefont
  {Baryakhtar}}, \bibinfo {author} {\bibfnamefont {M.}~\bibnamefont {Galanis}},
  \bibinfo {author} {\bibfnamefont {R.}~\bibnamefont {Lasenby}}, \ and\
  \bibinfo {author} {\bibfnamefont {O.}~\bibnamefont {Simon}},\ }\href
  {\doibase 10.1103/PhysRevD.103.095019} {\bibfield  {journal} {\bibinfo
  {journal} {Phys. Rev. D}\ }\textbf {\bibinfo {volume} {103}},\ \bibinfo
  {pages} {095019} (\bibinfo {year} {2021})},\ \Eprint
  {http://arxiv.org/abs/2011.11646} {arXiv:2011.11646 [hep-ph]} \BibitemShut
  {NoStop}%
\bibitem [{\citenamefont {Omiya}\ \emph {et~al.}(2021)\citenamefont {Omiya},
  \citenamefont {Takahashi},\ and\ \citenamefont {Tanaka}}]{Omiya:2020vji}%
  \BibitemOpen
  \bibfield  {author} {\bibinfo {author} {\bibfnamefont {H.}~\bibnamefont
  {Omiya}}, \bibinfo {author} {\bibfnamefont {T.}~\bibnamefont {Takahashi}}, \
  and\ \bibinfo {author} {\bibfnamefont {T.}~\bibnamefont {Tanaka}},\ }\href
  {\doibase 10.1093/ptep/ptab032} {\bibfield  {journal} {\bibinfo  {journal}
  {PTEP}\ }\textbf {\bibinfo {volume} {2021}},\ \bibinfo {pages} {043E02}
  (\bibinfo {year} {2021})},\ \Eprint {http://arxiv.org/abs/2012.03473}
  {arXiv:2012.03473 [gr-qc]} \BibitemShut {NoStop}%
\bibitem [{\citenamefont {Yoshino}\ and\ \citenamefont
  {Kodama}(2012)}]{Yoshino:2012kn}%
  \BibitemOpen
  \bibfield  {author} {\bibinfo {author} {\bibfnamefont {H.}~\bibnamefont
  {Yoshino}}\ and\ \bibinfo {author} {\bibfnamefont {H.}~\bibnamefont
  {Kodama}},\ }\href {\doibase 10.1143/PTP.128.153} {\bibfield  {journal}
  {\bibinfo  {journal} {Prog. Theor. Phys.}\ }\textbf {\bibinfo {volume}
  {128}},\ \bibinfo {pages} {153} (\bibinfo {year} {2012})},\ \Eprint
  {http://arxiv.org/abs/1203.5070} {arXiv:1203.5070 [gr-qc]} \BibitemShut
  {NoStop}%
\bibitem [{\citenamefont {Yoshino}\ and\ \citenamefont
  {Kodama}(2015{\natexlab{a}})}]{Yoshino:2015nsa}%
  \BibitemOpen
  \bibfield  {author} {\bibinfo {author} {\bibfnamefont {H.}~\bibnamefont
  {Yoshino}}\ and\ \bibinfo {author} {\bibfnamefont {H.}~\bibnamefont
  {Kodama}},\ }\href {\doibase 10.1088/0264-9381/32/21/214001} {\bibfield
  {journal} {\bibinfo  {journal} {Class. Quant. Grav.}\ }\textbf {\bibinfo
  {volume} {32}},\ \bibinfo {pages} {214001} (\bibinfo {year}
  {2015}{\natexlab{a}})},\ \Eprint {http://arxiv.org/abs/1505.00714}
  {arXiv:1505.00714 [gr-qc]} \BibitemShut {NoStop}%
\bibitem [{\citenamefont {Sen}(2018)}]{Sen:2018cjt}%
  \BibitemOpen
  \bibfield  {author} {\bibinfo {author} {\bibfnamefont {S.}~\bibnamefont
  {Sen}},\ }\href {\doibase 10.1103/PhysRevD.98.103012} {\bibfield  {journal}
  {\bibinfo  {journal} {Phys. Rev. D}\ }\textbf {\bibinfo {volume} {98}},\
  \bibinfo {pages} {103012} (\bibinfo {year} {2018})},\ \Eprint
  {http://arxiv.org/abs/1805.06471} {arXiv:1805.06471 [hep-ph]} \BibitemShut
  {NoStop}%
\bibitem [{\citenamefont {Blas}\ and\ \citenamefont
  {Witte}(2020{\natexlab{a}})}]{Blas:2020nbs}%
  \BibitemOpen
  \bibfield  {author} {\bibinfo {author} {\bibfnamefont {D.}~\bibnamefont
  {Blas}}\ and\ \bibinfo {author} {\bibfnamefont {S.~J.}\ \bibnamefont
  {Witte}},\ }\href {\doibase 10.1103/PhysRevD.102.103018} {\bibfield
  {journal} {\bibinfo  {journal} {Phys. Rev. D}\ }\textbf {\bibinfo {volume}
  {102}},\ \bibinfo {pages} {103018} (\bibinfo {year} {2020}{\natexlab{a}})},\
  \Eprint {http://arxiv.org/abs/2009.10074} {arXiv:2009.10074 [astro-ph.CO]}
  \BibitemShut {NoStop}%
\bibitem [{\citenamefont {Blas}\ and\ \citenamefont
  {Witte}(2020{\natexlab{b}})}]{Blas:2020kaa}%
  \BibitemOpen
  \bibfield  {author} {\bibinfo {author} {\bibfnamefont {D.}~\bibnamefont
  {Blas}}\ and\ \bibinfo {author} {\bibfnamefont {S.~J.}\ \bibnamefont
  {Witte}},\ }\href {\doibase 10.1103/PhysRevD.102.123018} {\bibfield
  {journal} {\bibinfo  {journal} {Phys. Rev. D}\ }\textbf {\bibinfo {volume}
  {102}},\ \bibinfo {pages} {123018} (\bibinfo {year} {2020}{\natexlab{b}})},\
  \Eprint {http://arxiv.org/abs/2009.10075} {arXiv:2009.10075 [hep-ph]}
  \BibitemShut {NoStop}%
\bibitem [{\citenamefont {Caputo}\ \emph {et~al.}(2021)\citenamefont {Caputo},
  \citenamefont {Witte}, \citenamefont {Blas},\ and\ \citenamefont
  {Pani}}]{Caputo:2021efm}%
  \BibitemOpen
  \bibfield  {author} {\bibinfo {author} {\bibfnamefont {A.}~\bibnamefont
  {Caputo}}, \bibinfo {author} {\bibfnamefont {S.~J.}\ \bibnamefont {Witte}},
  \bibinfo {author} {\bibfnamefont {D.}~\bibnamefont {Blas}}, \ and\ \bibinfo
  {author} {\bibfnamefont {P.}~\bibnamefont {Pani}},\ }\href@noop {} {\
  (\bibinfo {year} {2021})},\ \Eprint {http://arxiv.org/abs/2102.11280}
  {arXiv:2102.11280 [hep-ph]} \BibitemShut {NoStop}%
\bibitem [{\citenamefont {Franzin}\ \emph {et~al.}(2021)\citenamefont
  {Franzin}, \citenamefont {Liberati},\ and\ \citenamefont
  {Oi}}]{Franzin:2021kvj}%
  \BibitemOpen
  \bibfield  {author} {\bibinfo {author} {\bibfnamefont {E.}~\bibnamefont
  {Franzin}}, \bibinfo {author} {\bibfnamefont {S.}~\bibnamefont {Liberati}}, \
  and\ \bibinfo {author} {\bibfnamefont {M.}~\bibnamefont {Oi}},\ }\href
  {\doibase 10.1103/PhysRevD.103.104034} {\bibfield  {journal} {\bibinfo
  {journal} {Phys. Rev. D}\ }\textbf {\bibinfo {volume} {103}},\ \bibinfo
  {pages} {104034} (\bibinfo {year} {2021})},\ \Eprint
  {http://arxiv.org/abs/2102.03152} {arXiv:2102.03152 [gr-qc]} \BibitemShut
  {NoStop}%
\bibitem [{\citenamefont {Guo}\ \emph {et~al.}(2021)\citenamefont {Guo},
  \citenamefont {Yuan},\ and\ \citenamefont {Huang}}]{Guo:2021xao}%
  \BibitemOpen
  \bibfield  {author} {\bibinfo {author} {\bibfnamefont {R.-Z.}\ \bibnamefont
  {Guo}}, \bibinfo {author} {\bibfnamefont {C.}~\bibnamefont {Yuan}}, \ and\
  \bibinfo {author} {\bibfnamefont {Q.-G.}\ \bibnamefont {Huang}},\ }\href@noop
  {} {\  (\bibinfo {year} {2021})},\ \Eprint {http://arxiv.org/abs/2109.03376}
  {arXiv:2109.03376 [gr-qc]} \BibitemShut {NoStop}%
\bibitem [{\citenamefont {Conlon}\ and\ \citenamefont
  {Herdeiro}(2018)}]{Conlon:2017hhi}%
  \BibitemOpen
  \bibfield  {author} {\bibinfo {author} {\bibfnamefont {J.~P.}\ \bibnamefont
  {Conlon}}\ and\ \bibinfo {author} {\bibfnamefont {C.~A.}\ \bibnamefont
  {Herdeiro}},\ }\href {\doibase 10.1016/j.physletb.2018.02.073} {\bibfield
  {journal} {\bibinfo  {journal} {Phys. Lett. B}\ }\textbf {\bibinfo {volume}
  {780}},\ \bibinfo {pages} {169} (\bibinfo {year} {2018})},\ \Eprint
  {http://arxiv.org/abs/1701.02034} {arXiv:1701.02034 [astro-ph.HE]}
  \BibitemShut {NoStop}%
\bibitem [{\citenamefont {Dima}\ and\ \citenamefont
  {Barausse}(2020)}]{Dima:2020rzg}%
  \BibitemOpen
  \bibfield  {author} {\bibinfo {author} {\bibfnamefont {A.}~\bibnamefont
  {Dima}}\ and\ \bibinfo {author} {\bibfnamefont {E.}~\bibnamefont
  {Barausse}},\ }\href {\doibase 10.1088/1361-6382/ab9ce0} {\bibfield
  {journal} {\bibinfo  {journal} {Class. Quant. Grav.}\ }\textbf {\bibinfo
  {volume} {37}},\ \bibinfo {pages} {175006} (\bibinfo {year} {2020})},\
  \Eprint {http://arxiv.org/abs/2001.11484} {arXiv:2001.11484 [gr-qc]}
  \BibitemShut {NoStop}%
\bibitem [{\citenamefont {Schnitzeler}(2012)}]{Schnitzeler:2012jq}%
  \BibitemOpen
  \bibfield  {author} {\bibinfo {author} {\bibfnamefont {D.~H. F.~M.}\
  \bibnamefont {Schnitzeler}},\ }\href {\doibase
  10.1111/j.1365-2966.2012.21869.x} {\bibfield  {journal} {\bibinfo  {journal}
  {Mon. Not. Roy. Astron. Soc.}\ }\textbf {\bibinfo {volume} {427}},\ \bibinfo
  {pages} {664} (\bibinfo {year} {2012})},\ \Eprint
  {http://arxiv.org/abs/1208.3045} {arXiv:1208.3045 [astro-ph.GA]} \BibitemShut
  {NoStop}%
\bibitem [{\citenamefont {Cardoso}\ \emph {et~al.}(2021)\citenamefont
  {Cardoso}, \citenamefont {Guo}, \citenamefont {Macedo},\ and\ \citenamefont
  {Pani}}]{Cardoso:2020nst}%
  \BibitemOpen
  \bibfield  {author} {\bibinfo {author} {\bibfnamefont {V.}~\bibnamefont
  {Cardoso}}, \bibinfo {author} {\bibfnamefont {W.-D.}\ \bibnamefont {Guo}},
  \bibinfo {author} {\bibfnamefont {C.~F.~B.}\ \bibnamefont {Macedo}}, \ and\
  \bibinfo {author} {\bibfnamefont {P.}~\bibnamefont {Pani}},\ }\href {\doibase
  10.1093/mnras/stab404} {\bibfield  {journal} {\bibinfo  {journal} {Mon. Not.
  Roy. Astron. Soc.}\ }\textbf {\bibinfo {volume} {503}},\ \bibinfo {pages}
  {563} (\bibinfo {year} {2021})},\ \Eprint {http://arxiv.org/abs/2009.07287}
  {arXiv:2009.07287 [gr-qc]} \BibitemShut {NoStop}%
\bibitem [{\citenamefont {Cannizzaro}\ \emph
  {et~al.}(2021{\natexlab{a}})\citenamefont {Cannizzaro}, \citenamefont
  {Caputo}, \citenamefont {Sberna},\ and\ \citenamefont
  {Pani}}]{Cannizzaro:2020uap}%
  \BibitemOpen
  \bibfield  {author} {\bibinfo {author} {\bibfnamefont {E.}~\bibnamefont
  {Cannizzaro}}, \bibinfo {author} {\bibfnamefont {A.}~\bibnamefont {Caputo}},
  \bibinfo {author} {\bibfnamefont {L.}~\bibnamefont {Sberna}}, \ and\ \bibinfo
  {author} {\bibfnamefont {P.}~\bibnamefont {Pani}},\ }\href {\doibase
  10.1103/PhysRevD.103.124018} {\bibfield  {journal} {\bibinfo  {journal}
  {Phys. Rev. D}\ }\textbf {\bibinfo {volume} {103}},\ \bibinfo {pages}
  {124018} (\bibinfo {year} {2021}{\natexlab{a}})},\ \Eprint
  {http://arxiv.org/abs/2012.05114} {arXiv:2012.05114 [gr-qc]} \BibitemShut
  {NoStop}%
\bibitem [{\citenamefont {Cannizzaro}\ \emph
  {et~al.}(2021{\natexlab{b}})\citenamefont {Cannizzaro}, \citenamefont
  {Caputo}, \citenamefont {Sberna},\ and\ \citenamefont
  {Pani}}]{Cannizzaro:2021zbp}%
  \BibitemOpen
  \bibfield  {author} {\bibinfo {author} {\bibfnamefont {E.}~\bibnamefont
  {Cannizzaro}}, \bibinfo {author} {\bibfnamefont {A.}~\bibnamefont {Caputo}},
  \bibinfo {author} {\bibfnamefont {L.}~\bibnamefont {Sberna}}, \ and\ \bibinfo
  {author} {\bibfnamefont {P.}~\bibnamefont {Pani}},\ }\href@noop {} {\
  (\bibinfo {year} {2021}{\natexlab{b}})},\ \Eprint
  {http://arxiv.org/abs/2107.01174} {arXiv:2107.01174 [gr-qc]} \BibitemShut
  {NoStop}%
\bibitem [{\citenamefont {Cardoso}\ \emph {et~al.}(2013)\citenamefont
  {Cardoso}, \citenamefont {Carucci}, \citenamefont {Pani},\ and\ \citenamefont
  {Sotiriou}}]{Cardoso:2013opa}%
  \BibitemOpen
  \bibfield  {author} {\bibinfo {author} {\bibfnamefont {V.}~\bibnamefont
  {Cardoso}}, \bibinfo {author} {\bibfnamefont {I.~P.}\ \bibnamefont
  {Carucci}}, \bibinfo {author} {\bibfnamefont {P.}~\bibnamefont {Pani}}, \
  and\ \bibinfo {author} {\bibfnamefont {T.~P.}\ \bibnamefont {Sotiriou}},\
  }\href {\doibase 10.1103/PhysRevD.88.044056} {\bibfield  {journal} {\bibinfo
  {journal} {Phys. Rev. D}\ }\textbf {\bibinfo {volume} {88}},\ \bibinfo
  {pages} {044056} (\bibinfo {year} {2013})},\ \Eprint
  {http://arxiv.org/abs/1305.6936} {arXiv:1305.6936 [gr-qc]} \BibitemShut
  {NoStop}%
\bibitem [{\citenamefont {Payne}\ \emph {et~al.}(2021)\citenamefont {Payne},
  \citenamefont {Sun}, \citenamefont {Kremer}, \citenamefont {Lasky},\ and\
  \citenamefont {Thrane}}]{Payne:2021ahy}%
  \BibitemOpen
  \bibfield  {author} {\bibinfo {author} {\bibfnamefont {E.}~\bibnamefont
  {Payne}}, \bibinfo {author} {\bibfnamefont {L.}~\bibnamefont {Sun}}, \bibinfo
  {author} {\bibfnamefont {K.}~\bibnamefont {Kremer}}, \bibinfo {author}
  {\bibfnamefont {P.~D.}\ \bibnamefont {Lasky}}, \ and\ \bibinfo {author}
  {\bibfnamefont {E.}~\bibnamefont {Thrane}},\ }\href@noop {} {\  (\bibinfo
  {year} {2021})},\ \Eprint {http://arxiv.org/abs/2107.11730} {arXiv:2107.11730
  [gr-qc]} \BibitemShut {NoStop}%
\bibitem [{\citenamefont {Yoshino}\ and\ \citenamefont
  {Kodama}(2015{\natexlab{b}})}]{Yoshino:2014wwa}%
  \BibitemOpen
  \bibfield  {author} {\bibinfo {author} {\bibfnamefont {H.}~\bibnamefont
  {Yoshino}}\ and\ \bibinfo {author} {\bibfnamefont {H.}~\bibnamefont
  {Kodama}},\ }\href {\doibase 10.1093/ptep/ptv067} {\bibfield  {journal}
  {\bibinfo  {journal} {PTEP}\ }\textbf {\bibinfo {volume} {2015}},\ \bibinfo
  {pages} {061E01} (\bibinfo {year} {2015}{\natexlab{b}})},\ \Eprint
  {http://arxiv.org/abs/1407.2030} {arXiv:1407.2030 [gr-qc]} \BibitemShut
  {NoStop}%
\bibitem [{\citenamefont {Baryakhtar}\ \emph {et~al.}(2017)\citenamefont
  {Baryakhtar}, \citenamefont {Lasenby},\ and\ \citenamefont
  {Teo}}]{Baryakhtar:2017ngi}%
  \BibitemOpen
  \bibfield  {author} {\bibinfo {author} {\bibfnamefont {M.}~\bibnamefont
  {Baryakhtar}}, \bibinfo {author} {\bibfnamefont {R.}~\bibnamefont {Lasenby}},
  \ and\ \bibinfo {author} {\bibfnamefont {M.}~\bibnamefont {Teo}},\ }\href
  {\doibase 10.1103/PhysRevD.96.035019} {\bibfield  {journal} {\bibinfo
  {journal} {Phys. Rev. D}\ }\textbf {\bibinfo {volume} {96}},\ \bibinfo
  {pages} {035019} (\bibinfo {year} {2017})},\ \Eprint
  {http://arxiv.org/abs/1704.05081} {arXiv:1704.05081 [hep-ph]} \BibitemShut
  {NoStop}%
\bibitem [{\citenamefont {Ng}\ \emph {et~al.}(2021)\citenamefont {Ng},
  \citenamefont {Vitale}, \citenamefont {Hannuksela},\ and\ \citenamefont
  {Li}}]{Ng:2020ruv}%
  \BibitemOpen
  \bibfield  {author} {\bibinfo {author} {\bibfnamefont {K.~K.~Y.}\
  \bibnamefont {Ng}}, \bibinfo {author} {\bibfnamefont {S.}~\bibnamefont
  {Vitale}}, \bibinfo {author} {\bibfnamefont {O.~A.}\ \bibnamefont
  {Hannuksela}}, \ and\ \bibinfo {author} {\bibfnamefont {T.~G.~F.}\
  \bibnamefont {Li}},\ }\href {\doibase 10.1103/PhysRevLett.126.151102}
  {\bibfield  {journal} {\bibinfo  {journal} {Phys. Rev. Lett.}\ }\textbf
  {\bibinfo {volume} {126}},\ \bibinfo {pages} {151102} (\bibinfo {year}
  {2021})},\ \Eprint {http://arxiv.org/abs/2011.06010} {arXiv:2011.06010
  [gr-qc]} \BibitemShut {NoStop}%
\bibitem [{\citenamefont {Brito}\ \emph
  {et~al.}(2017{\natexlab{a}})\citenamefont {Brito}, \citenamefont {Ghosh},
  \citenamefont {Barausse}, \citenamefont {Berti}, \citenamefont {Cardoso},
  \citenamefont {Dvorkin}, \citenamefont {Klein},\ and\ \citenamefont
  {Pani}}]{Brito:2017wnc}%
  \BibitemOpen
  \bibfield  {author} {\bibinfo {author} {\bibfnamefont {R.}~\bibnamefont
  {Brito}}, \bibinfo {author} {\bibfnamefont {S.}~\bibnamefont {Ghosh}},
  \bibinfo {author} {\bibfnamefont {E.}~\bibnamefont {Barausse}}, \bibinfo
  {author} {\bibfnamefont {E.}~\bibnamefont {Berti}}, \bibinfo {author}
  {\bibfnamefont {V.}~\bibnamefont {Cardoso}}, \bibinfo {author} {\bibfnamefont
  {I.}~\bibnamefont {Dvorkin}}, \bibinfo {author} {\bibfnamefont
  {A.}~\bibnamefont {Klein}}, \ and\ \bibinfo {author} {\bibfnamefont
  {P.}~\bibnamefont {Pani}},\ }\href {\doibase 10.1103/PhysRevLett.119.131101}
  {\bibfield  {journal} {\bibinfo  {journal} {Phys. Rev. Lett.}\ }\textbf
  {\bibinfo {volume} {119}},\ \bibinfo {pages} {131101} (\bibinfo {year}
  {2017}{\natexlab{a}})},\ \Eprint {http://arxiv.org/abs/1706.05097}
  {arXiv:1706.05097 [gr-qc]} \BibitemShut {NoStop}%
\bibitem [{\citenamefont {Brito}\ \emph
  {et~al.}(2017{\natexlab{b}})\citenamefont {Brito}, \citenamefont {Ghosh},
  \citenamefont {Barausse}, \citenamefont {Berti}, \citenamefont {Cardoso},
  \citenamefont {Dvorkin}, \citenamefont {Klein},\ and\ \citenamefont
  {Pani}}]{Brito:2017zvb}%
  \BibitemOpen
  \bibfield  {author} {\bibinfo {author} {\bibfnamefont {R.}~\bibnamefont
  {Brito}}, \bibinfo {author} {\bibfnamefont {S.}~\bibnamefont {Ghosh}},
  \bibinfo {author} {\bibfnamefont {E.}~\bibnamefont {Barausse}}, \bibinfo
  {author} {\bibfnamefont {E.}~\bibnamefont {Berti}}, \bibinfo {author}
  {\bibfnamefont {V.}~\bibnamefont {Cardoso}}, \bibinfo {author} {\bibfnamefont
  {I.}~\bibnamefont {Dvorkin}}, \bibinfo {author} {\bibfnamefont
  {A.}~\bibnamefont {Klein}}, \ and\ \bibinfo {author} {\bibfnamefont
  {P.}~\bibnamefont {Pani}},\ }\href {\doibase 10.1103/PhysRevD.96.064050}
  {\bibfield  {journal} {\bibinfo  {journal} {Phys. Rev. D}\ }\textbf {\bibinfo
  {volume} {96}},\ \bibinfo {pages} {064050} (\bibinfo {year}
  {2017}{\natexlab{b}})},\ \Eprint {http://arxiv.org/abs/1706.06311}
  {arXiv:1706.06311 [gr-qc]} \BibitemShut {NoStop}%
\bibitem [{\citenamefont {Tsukada}\ \emph {et~al.}(2021)\citenamefont
  {Tsukada}, \citenamefont {Brito}, \citenamefont {East},\ and\ \citenamefont
  {Siemonsen}}]{Tsukada:2020lgt}%
  \BibitemOpen
  \bibfield  {author} {\bibinfo {author} {\bibfnamefont {L.}~\bibnamefont
  {Tsukada}}, \bibinfo {author} {\bibfnamefont {R.}~\bibnamefont {Brito}},
  \bibinfo {author} {\bibfnamefont {W.~E.}\ \bibnamefont {East}}, \ and\
  \bibinfo {author} {\bibfnamefont {N.}~\bibnamefont {Siemonsen}},\ }\href
  {\doibase 10.1103/PhysRevD.103.083005} {\bibfield  {journal} {\bibinfo
  {journal} {Phys. Rev. D}\ }\textbf {\bibinfo {volume} {103}},\ \bibinfo
  {pages} {083005} (\bibinfo {year} {2021})},\ \Eprint
  {http://arxiv.org/abs/2011.06995} {arXiv:2011.06995 [astro-ph.HE]}
  \BibitemShut {NoStop}%
\bibitem [{\citenamefont {Zhu}\ \emph {et~al.}(2020)\citenamefont {Zhu},
  \citenamefont {Baryakhtar}, \citenamefont {Papa}, \citenamefont {Tsuna},
  \citenamefont {Kawanaka},\ and\ \citenamefont {Eggenstein}}]{Zhu:2020tht}%
  \BibitemOpen
  \bibfield  {author} {\bibinfo {author} {\bibfnamefont {S.~J.}\ \bibnamefont
  {Zhu}}, \bibinfo {author} {\bibfnamefont {M.}~\bibnamefont {Baryakhtar}},
  \bibinfo {author} {\bibfnamefont {M.~A.}\ \bibnamefont {Papa}}, \bibinfo
  {author} {\bibfnamefont {D.}~\bibnamefont {Tsuna}}, \bibinfo {author}
  {\bibfnamefont {N.}~\bibnamefont {Kawanaka}}, \ and\ \bibinfo {author}
  {\bibfnamefont {H.-B.}\ \bibnamefont {Eggenstein}},\ }\href {\doibase
  10.1103/PhysRevD.102.063020} {\bibfield  {journal} {\bibinfo  {journal}
  {Phys. Rev. D}\ }\textbf {\bibinfo {volume} {102}},\ \bibinfo {pages}
  {063020} (\bibinfo {year} {2020})},\ \Eprint
  {http://arxiv.org/abs/2003.03359} {arXiv:2003.03359 [gr-qc]} \BibitemShut
  {NoStop}%
\bibitem [{\citenamefont {Cardoso}\ \emph {et~al.}(2017)\citenamefont
  {Cardoso}, \citenamefont {Pani},\ and\ \citenamefont {Yu}}]{Cardoso:2017kgn}%
  \BibitemOpen
  \bibfield  {author} {\bibinfo {author} {\bibfnamefont {V.}~\bibnamefont
  {Cardoso}}, \bibinfo {author} {\bibfnamefont {P.}~\bibnamefont {Pani}}, \
  and\ \bibinfo {author} {\bibfnamefont {T.-T.}\ \bibnamefont {Yu}},\ }\href
  {\doibase 10.1103/PhysRevD.95.124056} {\bibfield  {journal} {\bibinfo
  {journal} {Phys. Rev. D}\ }\textbf {\bibinfo {volume} {95}},\ \bibinfo
  {pages} {124056} (\bibinfo {year} {2017})},\ \Eprint
  {http://arxiv.org/abs/1704.06151} {arXiv:1704.06151 [gr-qc]} \BibitemShut
  {NoStop}%
\bibitem [{\citenamefont {Day}\ and\ \citenamefont
  {McDonald}(2019)}]{Day:2019bbh}%
  \BibitemOpen
  \bibfield  {author} {\bibinfo {author} {\bibfnamefont {F.~V.}\ \bibnamefont
  {Day}}\ and\ \bibinfo {author} {\bibfnamefont {J.~I.}\ \bibnamefont
  {McDonald}},\ }\href {\doibase 10.1088/1475-7516/2019/10/051} {\bibfield
  {journal} {\bibinfo  {journal} {JCAP}\ }\textbf {\bibinfo {volume} {10}},\
  \bibinfo {pages} {051} (\bibinfo {year} {2019})},\ \Eprint
  {http://arxiv.org/abs/1904.08341} {arXiv:1904.08341 [hep-ph]} \BibitemShut
  {NoStop}%
\bibitem [{\citenamefont {Pani}\ and\ \citenamefont
  {Loeb}(2013)}]{Pani:2013hpa}%
  \BibitemOpen
  \bibfield  {author} {\bibinfo {author} {\bibfnamefont {P.}~\bibnamefont
  {Pani}}\ and\ \bibinfo {author} {\bibfnamefont {A.}~\bibnamefont {Loeb}},\
  }\href {\doibase 10.1103/PhysRevD.88.041301} {\bibfield  {journal} {\bibinfo
  {journal} {Phys. Rev. D}\ }\textbf {\bibinfo {volume} {88}},\ \bibinfo
  {pages} {041301} (\bibinfo {year} {2013})},\ \Eprint
  {http://arxiv.org/abs/1307.5176} {arXiv:1307.5176 [astro-ph.CO]} \BibitemShut
  {NoStop}%
\bibitem [{\citenamefont {Lorimer}\ \emph {et~al.}(2007)\citenamefont
  {Lorimer}, \citenamefont {Bailes}, \citenamefont {McLaughlin}, \citenamefont
  {Narkevic},\ and\ \citenamefont {Crawford}}]{Lorimer:2007qn}%
  \BibitemOpen
  \bibfield  {author} {\bibinfo {author} {\bibfnamefont {D.}~\bibnamefont
  {Lorimer}}, \bibinfo {author} {\bibfnamefont {M.}~\bibnamefont {Bailes}},
  \bibinfo {author} {\bibfnamefont {M.}~\bibnamefont {McLaughlin}}, \bibinfo
  {author} {\bibfnamefont {D.}~\bibnamefont {Narkevic}}, \ and\ \bibinfo
  {author} {\bibfnamefont {F.}~\bibnamefont {Crawford}},\ }\href {\doibase
  10.1126/science.1147532} {\bibfield  {journal} {\bibinfo  {journal}
  {Science}\ }\textbf {\bibinfo {volume} {318}},\ \bibinfo {pages} {777}
  (\bibinfo {year} {2007})},\ \Eprint {http://arxiv.org/abs/0709.4301}
  {arXiv:0709.4301 [astro-ph]} \BibitemShut {NoStop}%
\bibitem [{\citenamefont {Katz}(2016)}]{Katz:2016dti}%
  \BibitemOpen
  \bibfield  {author} {\bibinfo {author} {\bibfnamefont {J.}~\bibnamefont
  {Katz}},\ }\href {\doibase 10.1142/S0217732316300135} {\bibfield  {journal}
  {\bibinfo  {journal} {Mod. Phys. Lett. A}\ }\textbf {\bibinfo {volume}
  {31}},\ \bibinfo {pages} {1630013} (\bibinfo {year} {2016})},\ \Eprint
  {http://arxiv.org/abs/1604.01799} {arXiv:1604.01799 [astro-ph.HE]}
  \BibitemShut {NoStop}%
\bibitem [{\citenamefont {Houde}\ \emph {et~al.}(2019)\citenamefont {Houde},
  \citenamefont {Rajabi}, \citenamefont {Gaensler}, \citenamefont {Mathews},\
  and\ \citenamefont {Tranchant}}]{Houde:2018yos}%
  \BibitemOpen
  \bibfield  {author} {\bibinfo {author} {\bibfnamefont {M.}~\bibnamefont
  {Houde}}, \bibinfo {author} {\bibfnamefont {F.}~\bibnamefont {Rajabi}},
  \bibinfo {author} {\bibfnamefont {B.~M.}\ \bibnamefont {Gaensler}}, \bibinfo
  {author} {\bibfnamefont {A.}~\bibnamefont {Mathews}}, \ and\ \bibinfo
  {author} {\bibfnamefont {V.}~\bibnamefont {Tranchant}},\ }\href {\doibase
  10.1093/mnras/sty3046} {\bibfield  {journal} {\bibinfo  {journal} {Mon. Not.
  Roy. Astron. Soc.}\ }\textbf {\bibinfo {volume} {482}},\ \bibinfo {pages}
  {5492} (\bibinfo {year} {2019})},\ \Eprint {http://arxiv.org/abs/1810.04364}
  {arXiv:1810.04364 [astro-ph.HE]} \BibitemShut {NoStop}%
\bibitem [{\citenamefont {Dolan}(2013{\natexlab{b}})}]{Dolan:2013}%
  \BibitemOpen
  \bibfield  {author} {\bibinfo {author} {\bibfnamefont {S.~R.}\ \bibnamefont
  {Dolan}},\ }\href {\doibase 10.1103/physrevd.87.124026} {\bibfield  {journal}
  {\bibinfo  {journal} {Physical Review D}\ }\textbf {\bibinfo {volume} {87}}
  (\bibinfo {year} {2013}{\natexlab{b}}),\
  10.1103/physrevd.87.124026}\BibitemShut {NoStop}%
\bibitem [{\citenamefont {Visser}(2008)}]{Visser:2007fj}%
  \BibitemOpen
  \bibfield  {author} {\bibinfo {author} {\bibfnamefont {M.}~\bibnamefont
  {Visser}},\ }\href@noop {} {\enquote {\bibinfo {title} {The kerr spacetime: A
  brief introduction},}\ } (\bibinfo {year} {2008}),\ \Eprint
  {http://arxiv.org/abs/0706.0622} {arXiv:0706.0622 [gr-qc]} \BibitemShut
  {NoStop}%
\bibitem [{\citenamefont {Hilditch}(2013)}]{Hilditch:2013sba}%
  \BibitemOpen
  \bibfield  {author} {\bibinfo {author} {\bibfnamefont {D.}~\bibnamefont
  {Hilditch}},\ }\href {\doibase 10.1142/S0217751X13400150} {\bibfield
  {journal} {\bibinfo  {journal} {Int. J. Mod. Phys. A}\ }\textbf {\bibinfo
  {volume} {28}},\ \bibinfo {pages} {1340015} (\bibinfo {year} {2013})},\
  \Eprint {http://arxiv.org/abs/1309.2012} {arXiv:1309.2012 [gr-qc]}
  \BibitemShut {NoStop}%
\bibitem [{\citenamefont {Palenzuela}\ \emph {et~al.}(2010)\citenamefont
  {Palenzuela}, \citenamefont {Lehner},\ and\ \citenamefont
  {Yoshida}}]{Palenzuela:2009hx}%
  \BibitemOpen
  \bibfield  {author} {\bibinfo {author} {\bibfnamefont {C.}~\bibnamefont
  {Palenzuela}}, \bibinfo {author} {\bibfnamefont {L.}~\bibnamefont {Lehner}},
  \ and\ \bibinfo {author} {\bibfnamefont {S.}~\bibnamefont {Yoshida}},\ }\href
  {\doibase 10.1103/PhysRevD.81.084007} {\bibfield  {journal} {\bibinfo
  {journal} {Phys. Rev. D}\ }\textbf {\bibinfo {volume} {81}},\ \bibinfo
  {pages} {084007} (\bibinfo {year} {2010})},\ \Eprint
  {http://arxiv.org/abs/0911.3889} {arXiv:0911.3889 [gr-qc]} \BibitemShut
  {NoStop}%
\bibitem [{\citenamefont {Clough}(2021)}]{Clough:2021qlv}%
  \BibitemOpen
  \bibfield  {author} {\bibinfo {author} {\bibfnamefont {K.}~\bibnamefont
  {Clough}},\ }\href@noop {} {\  (\bibinfo {year} {2021})},\ \Eprint
  {http://arxiv.org/abs/2104.13420} {arXiv:2104.13420 [gr-qc]} \BibitemShut
  {NoStop}%
\bibitem [{\citenamefont {{Bondi}}(1952)}]{Bondi_accretion}%
  \BibitemOpen
  \bibfield  {author} {\bibinfo {author} {\bibfnamefont {H.}~\bibnamefont
  {{Bondi}}},\ }\href {\doibase 10.1093/mnras/112.2.195} {\bibfield  {journal}
  {\bibinfo  {journal} {MNRAS}\ }\textbf {\bibinfo {volume} {112}},\ \bibinfo
  {pages} {195} (\bibinfo {year} {1952})}\BibitemShut {NoStop}%
\bibitem [{\citenamefont {Narayan}\ and\ \citenamefont {Yi}(1994)}]{ADAF_1994}%
  \BibitemOpen
  \bibfield  {author} {\bibinfo {author} {\bibfnamefont {R.}~\bibnamefont
  {Narayan}}\ and\ \bibinfo {author} {\bibfnamefont {I.}~\bibnamefont {Yi}},\
  }\href {\doibase 10.1086/187381} {\bibfield  {journal} {\bibinfo  {journal}
  {The Astrophysical Journal}\ }\textbf {\bibinfo {volume} {428}},\ \bibinfo
  {pages} {L13} (\bibinfo {year} {1994})}\BibitemShut {NoStop}%
\bibitem [{\citenamefont {Narayan}\ and\ \citenamefont
  {Yi}(1995{\natexlab{a}})}]{ADAF_1995}%
  \BibitemOpen
  \bibfield  {author} {\bibinfo {author} {\bibfnamefont {R.}~\bibnamefont
  {Narayan}}\ and\ \bibinfo {author} {\bibfnamefont {I.}~\bibnamefont {Yi}},\
  }\href {\doibase 10.1086/176343} {\bibfield  {journal} {\bibinfo  {journal}
  {The Astrophysical Journal}\ }\textbf {\bibinfo {volume} {452}},\ \bibinfo
  {pages} {710} (\bibinfo {year} {1995}{\natexlab{a}})}\BibitemShut {NoStop}%
\bibitem [{\citenamefont {Narayan}\ and\ \citenamefont
  {Yi}(1995{\natexlab{b}})}]{ADAF_1995_2}%
  \BibitemOpen
  \bibfield  {author} {\bibinfo {author} {\bibfnamefont {R.}~\bibnamefont
  {Narayan}}\ and\ \bibinfo {author} {\bibfnamefont {I.}~\bibnamefont {Yi}},\
  }\href {\doibase 10.1086/175599} {\bibfield  {journal} {\bibinfo  {journal}
  {The Astrophysical Journal}\ }\textbf {\bibinfo {volume} {444}},\ \bibinfo
  {pages} {231} (\bibinfo {year} {1995}{\natexlab{b}})}\BibitemShut {NoStop}%
\bibitem [{\citenamefont {Stanzione}\ \emph {et~al.}(2020)\citenamefont
  {Stanzione}, \citenamefont {West}, \citenamefont {Evans}, \citenamefont
  {Minyard}, \citenamefont {Ghattas},\ and\ \citenamefont
  {Panda}}]{10.1145/3311790.3396656}%
  \BibitemOpen
  \bibfield  {author} {\bibinfo {author} {\bibfnamefont {D.}~\bibnamefont
  {Stanzione}}, \bibinfo {author} {\bibfnamefont {J.}~\bibnamefont {West}},
  \bibinfo {author} {\bibfnamefont {R.~T.}\ \bibnamefont {Evans}}, \bibinfo
  {author} {\bibfnamefont {T.}~\bibnamefont {Minyard}}, \bibinfo {author}
  {\bibfnamefont {O.}~\bibnamefont {Ghattas}}, \ and\ \bibinfo {author}
  {\bibfnamefont {D.~K.}\ \bibnamefont {Panda}},\ }in\ \href {\doibase
  10.1145/3311790.3396656} {\emph {\bibinfo {booktitle} {Practice and
  Experience in Advanced Research Computing}}},\ \bibinfo {series and number}
  {PEARC '20}\ (\bibinfo  {publisher} {Association for Computing Machinery},\
  \bibinfo {address} {New York, NY, USA},\ \bibinfo {year} {2020})\ p.\
  \bibinfo {pages} {106–111}\BibitemShut {NoStop}%
\bibitem [{\citenamefont {Clough}\ \emph {et~al.}(2015)\citenamefont {Clough},
  \citenamefont {Figueras}, \citenamefont {Finkel}, \citenamefont {Kunesch},
  \citenamefont {Lim},\ and\ \citenamefont {Tunyasuvunakool}}]{Clough:2015sqa}%
  \BibitemOpen
  \bibfield  {author} {\bibinfo {author} {\bibfnamefont {K.}~\bibnamefont
  {Clough}}, \bibinfo {author} {\bibfnamefont {P.}~\bibnamefont {Figueras}},
  \bibinfo {author} {\bibfnamefont {H.}~\bibnamefont {Finkel}}, \bibinfo
  {author} {\bibfnamefont {M.}~\bibnamefont {Kunesch}}, \bibinfo {author}
  {\bibfnamefont {E.~A.}\ \bibnamefont {Lim}}, \ and\ \bibinfo {author}
  {\bibfnamefont {S.}~\bibnamefont {Tunyasuvunakool}},\ }\href {\doibase
  10.1088/0264-9381/32/24/245011} {\bibfield  {journal} {\bibinfo  {journal}
  {Class. Quant. Grav.}\ }\textbf {\bibinfo {volume} {32}},\ \bibinfo {pages}
  {24} (\bibinfo {year} {2015})},\ \Eprint {http://arxiv.org/abs/1503.03436}
  {arXiv:1503.03436 [gr-qc]} \BibitemShut {NoStop}%
\bibitem [{\citenamefont {Andrade}\ \emph {et~al.}(2021)\citenamefont
  {Andrade}, \citenamefont {Salo}, \citenamefont {Aurrekoetxea}, \citenamefont
  {Bamber}, \citenamefont {Clough}, \citenamefont {Croft}, \citenamefont
  {de~Jong}, \citenamefont {Drew}, \citenamefont {Duran}, \citenamefont
  {Ferreira}, \citenamefont {Figueras}, \citenamefont {Finkel}, \citenamefont
  {Fran\c{c}a}, \citenamefont {Ge}, \citenamefont {Gu}, \citenamefont {Helfer},
  \citenamefont {Jäykkä}, \citenamefont {Joana}, \citenamefont {Kunesch},
  \citenamefont {Kornet}, \citenamefont {Lim}, \citenamefont {Muia},
  \citenamefont {Nazari}, \citenamefont {Radia}, \citenamefont {Ripley},
  \citenamefont {Shellard}, \citenamefont {Sperhake}, \citenamefont {Traykova},
  \citenamefont {Tunyasuvunakool}, \citenamefont {Wang}, \citenamefont
  {Widdicombe},\ and\ \citenamefont {Wong}}]{Andrade2021}%
  \BibitemOpen
  \bibfield  {author} {\bibinfo {author} {\bibfnamefont {T.}~\bibnamefont
  {Andrade}}, \bibinfo {author} {\bibfnamefont {L.~A.}\ \bibnamefont {Salo}},
  \bibinfo {author} {\bibfnamefont {J.~C.}\ \bibnamefont {Aurrekoetxea}},
  \bibinfo {author} {\bibfnamefont {J.}~\bibnamefont {Bamber}}, \bibinfo
  {author} {\bibfnamefont {K.}~\bibnamefont {Clough}}, \bibinfo {author}
  {\bibfnamefont {R.}~\bibnamefont {Croft}}, \bibinfo {author} {\bibfnamefont
  {E.}~\bibnamefont {de~Jong}}, \bibinfo {author} {\bibfnamefont
  {A.}~\bibnamefont {Drew}}, \bibinfo {author} {\bibfnamefont {A.}~\bibnamefont
  {Duran}}, \bibinfo {author} {\bibfnamefont {P.~G.}\ \bibnamefont {Ferreira}},
  \bibinfo {author} {\bibfnamefont {P.}~\bibnamefont {Figueras}}, \bibinfo
  {author} {\bibfnamefont {H.}~\bibnamefont {Finkel}}, \bibinfo {author}
  {\bibfnamefont {T.}~\bibnamefont {Fran\c{c}a}}, \bibinfo {author}
  {\bibfnamefont {B.-X.}\ \bibnamefont {Ge}}, \bibinfo {author} {\bibfnamefont
  {C.}~\bibnamefont {Gu}}, \bibinfo {author} {\bibfnamefont {T.}~\bibnamefont
  {Helfer}}, \bibinfo {author} {\bibfnamefont {J.}~\bibnamefont {Jäykkä}},
  \bibinfo {author} {\bibfnamefont {C.}~\bibnamefont {Joana}}, \bibinfo
  {author} {\bibfnamefont {M.}~\bibnamefont {Kunesch}}, \bibinfo {author}
  {\bibfnamefont {K.}~\bibnamefont {Kornet}}, \bibinfo {author} {\bibfnamefont
  {E.~A.}\ \bibnamefont {Lim}}, \bibinfo {author} {\bibfnamefont
  {F.}~\bibnamefont {Muia}}, \bibinfo {author} {\bibfnamefont {Z.}~\bibnamefont
  {Nazari}}, \bibinfo {author} {\bibfnamefont {M.}~\bibnamefont {Radia}},
  \bibinfo {author} {\bibfnamefont {J.}~\bibnamefont {Ripley}}, \bibinfo
  {author} {\bibfnamefont {P.}~\bibnamefont {Shellard}}, \bibinfo {author}
  {\bibfnamefont {U.}~\bibnamefont {Sperhake}}, \bibinfo {author}
  {\bibfnamefont {D.}~\bibnamefont {Traykova}}, \bibinfo {author}
  {\bibfnamefont {S.}~\bibnamefont {Tunyasuvunakool}}, \bibinfo {author}
  {\bibfnamefont {Z.}~\bibnamefont {Wang}}, \bibinfo {author} {\bibfnamefont
  {J.~Y.}\ \bibnamefont {Widdicombe}}, \ and\ \bibinfo {author} {\bibfnamefont
  {K.}~\bibnamefont {Wong}},\ }\href {\doibase 10.21105/joss.03703} {\bibfield
  {journal} {\bibinfo  {journal} {Journal of Open Source Software}\ }\textbf
  {\bibinfo {volume} {6}},\ \bibinfo {pages} {3703} (\bibinfo {year}
  {2021})}\BibitemShut {NoStop}%
\bibitem [{\citenamefont {Radia}\ \emph {et~al.}(2021)\citenamefont {Radia},
  \citenamefont {Sperhake}, \citenamefont {Drew}, \citenamefont {Clough},
  \citenamefont {Lim}, \citenamefont {Ripley}, \citenamefont {Aurrekoetxea},
  \citenamefont {Fran\c{c}a},\ and\ \citenamefont {Helfer}}]{Radia:2021smk}%
  \BibitemOpen
  \bibfield  {author} {\bibinfo {author} {\bibfnamefont {M.}~\bibnamefont
  {Radia}}, \bibinfo {author} {\bibfnamefont {U.}~\bibnamefont {Sperhake}},
  \bibinfo {author} {\bibfnamefont {A.}~\bibnamefont {Drew}}, \bibinfo {author}
  {\bibfnamefont {K.}~\bibnamefont {Clough}}, \bibinfo {author} {\bibfnamefont
  {E.~A.}\ \bibnamefont {Lim}}, \bibinfo {author} {\bibfnamefont {J.~L.}\
  \bibnamefont {Ripley}}, \bibinfo {author} {\bibfnamefont {J.~C.}\
  \bibnamefont {Aurrekoetxea}}, \bibinfo {author} {\bibfnamefont
  {T.}~\bibnamefont {Fran\c{c}a}}, \ and\ \bibinfo {author} {\bibfnamefont
  {T.}~\bibnamefont {Helfer}},\ }\href@noop {} {\  (\bibinfo {year} {2021})},\
  \Eprint {http://arxiv.org/abs/2112.10567} {arXiv:2112.10567 [gr-qc]}
  \BibitemShut {NoStop}%
\end{thebibliography}%

\end{document}